\documentstyle[preprint,aps,prl]{revtex}

\begin{document}
\draft

\preprint{To be published in: {\it Molecular Electronics}, ed.\ by J.\ Jortner 
	 and M.\ A.\ Ratner (Blackwell, Oxford)}

\title{Coulomb Blockade and Digital Single-Electron Devices}

\author{Alexander N. Korotkov}
\address{Department of Physics, State University of New York, Stony 
Brook, NY 11794-3800 \\ and \\ 
Nuclear Physics Institute, Moscow State University, Moscow 119899, 
Russia}


\maketitle

\begin{abstract}
        Tunneling of single electrons has been thoroughly studied both
theoretically and experimentally during last ten years. By the present
time the basic physics is well understood, and 
creation of useful single-electron devices becomes the important issue.
Single-electron tunneling seems to be the most promising candidate
to be used in the future integrated digital circuits with the typical
size scale of few nanometers and below, i.e.\ in the molecular
electronics. 
        In the review we first briefly discuss the physics of 
single-electron tunneling and the operation of the single-electron
transistor. After that, we concentrate on the hypothetical ultradense
digital single-electron circuits and discuss the different proposed
families of them. The last part of the review considers the issues
of the discrete energy spectrum and the finite tunnel barrier height
which are important for the molecular-size single-electron devices.

\end{abstract}

	\newpage
        \section[introduction]{Introduction}    

        The continuous trend in scaling down the component size 
leads the integrated electronics into the domain of so-called 
``mesoscopics'', the dimension area between the microscopic 
and macroscopic worlds. The Coulomb blockade and other single-electron 
charging effects (for other reviews see, e.g., Refs.\
\onlinecite{Av-Likh-mes,Likh-IBM,SCT,Likh-SA,Kastner}) 
are among the most prominent phenomena of mesoscopic physics. Despite these 
effects are caused by single electrons that makes a connection with 
microscopic world, they are usually well-described in terms of 
macroscopic ``electrical-engineering'' quantities like capacitances 
and resistances.
        Single-electron effects will play an important role in almost
any electronic device with dimensions  below $\sim$30 nm. Moreover,
they can be used as the new physical basis for the operation of
nanoscale digital circuits.

        The main component of single-electronics is a tunnel junction 
with a very small capacitance. These junctions can be implemented 
using a variety 
of materials: metal-insulator-metal structures, GaAs quantum dots, 
silicon structures, large molecules with conducting cores, etc. If the 
size (and, hence, the effective electrical capacitance $C$) of the 
junction is sufficiently small, then the tunneling of only one 
electron may produce a noticeable change $e/C$ of the voltage across 
the junction. The discreteness of this change which is a consequence 
of the electrical charge discreteness, leads to a number of effects 
which constitute the field of single-electronics.
        The most widely known effect is the Coulomb blockade. This is
the suppression of tunneling at voltages $|V|<e/2C$ because in this case 
the tunneling 
would increase the electrostatic energy of the capacitor:
$C(V\pm e/C)^2/2>CV^2/2$.
Besides the ``Coulomb blockade'' and ``single-electronics'', the other
key words of the field are ``correlated tunneling'' and 
``single charge tunneling''.

        The typical capacitance in most present day experiments is
on the order of $10^{-16}$ F (the simplest well-established technology 
uses metal junctions with an area about 50$\times$50 nm$^2$). It corresponds 
to the voltage scale $e/C$ on the order of one millivolt that is 
already sufficiently large to allow experimental studies. To avoid 
the smearing of single-electron effects by thermal fluctuations, the 
thermal energy $k_BT$ should be much less then the typical one-electron 
charging energy,
\begin{equation}
        k_B T << e^2/2C.
\label{T<<}\end{equation} 
        For $C\approx 10^{-16}$ F this condition limits the temperature 
to $T \alt 1$ K. 
        As a consequence, at the present stage of development, the 
practical use of single-electron devices is limited to scientific
experimentation and fundamental metrology.
        To achieve the real industrial impact, the operation temperature must
be increased up to 300 K (or at least to 77 K) that requires a dramatic 
decrease of the typical size of the components.
        There is already a considerable number of experiments in which
the room temperature or liquid nitrogen temperature operation of 
simple single-electron ``devices'' was reported, but  these
systems are not well-controllable and reproducible so far.

        Capacitances as small as $10^{-19}$ F can be achieved in
principle in the ``molecular electronic devices'' using conducting
clusters of atoms (with diameter of 1 nm or even less) embedded in a
molecular matrix. In this case the energy scale $e^2/2C$ would be about 
one electron-volt, and the room temperature operation
could be ensured. Notice that at the size scale below few nanometers, 
single-electronics
gradually enters the areas of chemistry and atomic physics.
Here, the discreteness of energy levels in conducting ``islands'' 
becomes an important factor, the electric capacitance becomes a not 
well-defined quantity, and the Coulomb blockade energy gradually 
converts into the energy of ionization 
and electron affinity. 
	However, the ideas of single-electronics are 
applicable even at this size scale, opening the possibility of 
information processing in molecular-based devices.

	\newpage
        \section[history]{ Brief history}

        The influence of the single-electron effects on the
conductance of thin granular metallic film was understood at least by 
Gorter \cite{Gorter} as early as in 1951. During the next 20 years
important contributions to this field came from the papers by
Neugebauer and Webb \cite{Neugebauer}, Giaver and Zeller 
\cite{Giaver}, and Lambe and Jaklevic \cite{Lambe}. The first
quantitative theory for a two-junction system was developed in 1975
by Kulik and Shekhter \cite{Kulik}.

        In the present-day meaning the single-electronics was
launched in the mid-1980s when the detailed theory of correlated
tunneling (now known as the ``orthodox'' theory) developed by Averin
and Likharev \cite{Av-Likh-86,Likh-87} was almost immediately
supported by first experiments with single-electron transistors
\cite{Fulton-D,Kuzmin-Likh}. Since that time there was a constantly 
growing interest both in theoretical and experimental single-electronics,
so that the total number of publications by the present time is 
well above a thousand.

        Single-electron effects were studied experimentally using
various materials and technologies. The oldest and most
developed technique is the fabrication of small tunnel junctions
by overlapping narrow strips of metal films using electron beam patterning 
and double-angle evaporation (see, e.g., Refs.\ \cite{Fulton-D,Delsing,%
Geerligs,Pothier-pump,Fulton-trap,Tuominen,Martinis-noise,LaFarge-trap,%
Pekola-PRL,Pashkin,Verbrugh,Tsai,Wahlgren}).
 Among the best achievements towards possible applications, 
let us mention single-electron transistors with the charge sensitivity
better than $10^{-4} e/\sqrt{\mbox{Hz}}$ (at 10 Hz) \cite{Visscher}, 
low-temperature
absolute thermometer with $\sim 1 \%$ accuracy \cite{Pekola}, the prototype
of the dc current standard with the relative accuracy better than 
$10^{-6}$ \cite{Martinis-pump}, and the
single-electron trap (``memory cell'') with the retention time more
than 12 hours \cite{Likh-trap}.
        
        Since 1990 \cite{Meirav} single-electron charging effects
are also studied in tunneling through small islands (``quantum
dots'') of 2D electron gas in GaAs-based heterostructures. It was predicted
\cite{Kor-dots} that in these structures 
the coexistence of the energy and charge quantizations should play
a much more important role than in metal
islands of comparable size. This fact was soon confirmed experimentally 
\cite{Goldman,McEuen}
(now the structures exhibiting both types of quantization are
sometimes called ``artificial atoms'' \cite{Kastner}). 
        The study of single-electron effects in the systems of quantum
dots was so far of mainly scientific interest (see, e.g., the latest 
papers \cite{Vaart,Duruoz,Waugh,Kouwenhoven,Klein}), 
and their technology is still too far from possible applications.
        As an exception let us mention the experiment \cite{Nakazato-mem}
in which the side-gated constrictions in $\delta$-doped GaAs were 
used to demonstrate the operation of the few-electron ``memory cell'' 
at 4 K. The Coulomb barrier in this case appears due to tunneling 
between randomly positioned small conducting islands in the constriction.

        There is a considerable number of experiments in which single
electrons tunnel from the tip of scanning tunneling microscope to the
substrate via a small conducting particle (see, e.g.\ Refs.\ 
\cite{vanBentum,Wilkins,Wan,Hanna,Schonenberger,Dorogi,Nejoh1,Nejoh2,%
Fischer,Zubilov,Dubois}).  The effect may survive up to 
the room temperature for sufficiently small metal particles 
\cite{Schonenberger,Dorogi}, and it can be even stronger when the tunneling 
via single molecules is studied \cite{Nejoh1,Nejoh2,Fischer,Zubilov,Dubois}.
        Scanning tunneling microscope can be also used for the fabrication
of the single-electron circuits \cite{Matsumoto}.

        Quite promising results were obtained recently in the experiments 
with silicon-based structures 
\cite{Yano,Takahashi,Fujiwara,Matsuoka,Chou}. The memory
effects and the operation of single-electron transistor were reported
even at room temperature. Taking into account the huge experience
accumulated in silicon technologies, the silicon-based devices
could be a real way to the integrated single-electronics.

        Ultradense integrated circuits is the most intriguing goal
of the single-electronics. There were many theoretical suggestions
on this topic. Let us mention the digital circuits based on 
single-electron transistors \cite{Likh-87,Tucker,Kor-trans,Chen}, 
the logic which uses single electrons to represent logic
bits \cite{Likh-Sem,Likh-Polon,Av-Likh-log,Naz-Vysh,Nakazato-log,Ancona}
(including various ``wireless'' logics \cite{Kor-isl,Param}) and 
background-charge-independent devices \cite{q0ind}.
        The practical realization of the integrated
single-electron circuits remains quite a questionable issue
because very serious technological
problems should be solved to reach this goal. Nevertheless,
the rapid progress in experimental single-electronics during a few last
years combined with the rapid improvement of the nanotechnology
makes room temperature single-electronics a 
candidate for the next generation of ultradense digital devices.


	\newpage
       \section{``Orthodox'' theory of single-electronics}

        The main object of the single-electronics is a small 
tunnel junction. The simplest approach described below works
very well for metallic junctions.
        Several specific
features of semiconductor and molecular-level systems will be
considered in Section VI.

        The  tunnel junction consists of two  electrodes
separated by an insulating layer and naturally has some electric 
capacitance $C$ depending on the geometry (in the simple case of the plain
capacitor $C=\varepsilon \varepsilon_0 S/d$ where $S$ is the area, $d$ is the
insulator thickness, and $\varepsilon$ is its dielectric constant). In 
contrast to the usual capacitor, in the tunnel junction electrons can 
pass through a sufficiently thin barrier (typically several nanometers).
        Let us assume the linear $I-V$ curve (in the absence of single-electron
effects), $I=V/R$, that is the typical case for metallic systems.
        
        The ``orthodox'' single-electronics \cite{Av-Likh-mes} deals 
with junctions having sufficiently large resistances,
\begin{equation}
        R >> R_Q=\pi \hbar /2e^2 \approx 6.4 k\Omega .
\label{R>>RQ} \end{equation}
        To understand the physical meaning of this condition let us
notice that the rate of tunneling in a junction biased by some voltage
$V$ is $\Gamma = V/eR$ so that the typical time between tunneling events 
is $1/\Gamma =eR/V$. The ``duration'' of a tunneling event due to
the uncertainty principle is $\hbar /eV$ (there are several other
definitions of the tunneling time -- see, e.g.\ Ref.\ \cite{Hauge}, 
however, they are not relevant to 
this problem). Hence, Eq.\ (\ref{R>>RQ})
simply means that the tunneling events do not overlap, and we can speak
about the separate tunneling of single electrons.

        In the case of small capacitance $C$ the voltage $V=V_b$ before 
the tunneling event is considerably different from the voltage 
$V_a=V_b-e/C$ after the event. Hence, it is not
clear which value should be used for the calculation of the tunneling 
rate. The simple guess is that we can take the average value as the 
effective voltage 
\begin{equation}
        V_{eff}=\frac{V_b+V_a}{2}=V_b-\frac{e}{2C}.
\label{Veff}\end{equation}
        This guess coincides with the result of the ``orthodox'' theory.
In fact, the effective voltage should be related to the change $W$ of
the electrostatic energy of the system (energy gain due to tunneling). 
In the case of a single capacitor charged initially by $Q=CV$ this 
change will be 
\begin{equation}
        W=\frac{Q^2}{2C}-\frac{(Q-e)^2}{2C}=e(V-\frac{e}{2C})=eV_{eff},
\label{W}\end{equation}
that coincides with Eq.\ (\ref{Veff}). This derivation is still valid if we
consider the tunnel junction being a part of the complex circuit.
The only difference is that we need to use the effective (total) 
capacitance of the junction calculated with account of the rest of
the circuit.

        For a given energy gain $W$ the tunneling rate in ``orthodox'' 
theory is calculated using the formula
\begin{equation}
        \Gamma = \frac{W}{e^2R(1-\exp (-W/k_B T))}
\label{rate}\end{equation}
where $T$ is the temperature. In case of zero temperature this expression 
transforms to $\Gamma =V_{eff}/eR$ for positive $W=eV_{eff}$ and
$\Gamma=0$ for negative $W$.

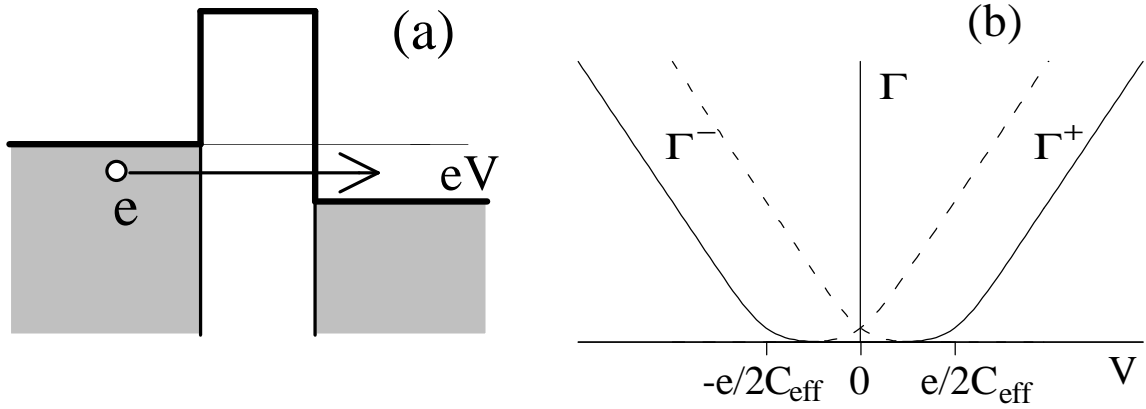
\begin{figure}
\caption{(a) Schematic energy diagram of the tunnel junction and (b)
the tunneling rates $\Gamma^+$ and $\Gamma^-$ for both directions
as functions of the voltage $V$ across the junction. The Coulomb
blockade suppresses the tunneling at $|V|<e/2C_{eff}$ (solid lines). 
The cusps of the curves 
are rounded due to finite temperature. The dashed lines show
$\Gamma^+$ and $\Gamma^-$ for the case without single-electron
effects ($e/C_{eff}=0$).}
\label{1j}\end{figure}

        Absence of tunneling for $W<0$ is natural because the processes 
which increase the free energy are forbidden by the second principle of
thermodynamics.
        Using Eq.\ (\ref{W}) we see that $W<0$ 
if the voltage across the junction is less than the threshold value 
$V_t=e/2C_{eff}$
that corresponds to the charge $Q=e/2$. This is the condition of the Coulomb 
blockade of tunneling (see Fig.\ \ref{1j}). For finite temperature the 
tunneling inside the blockade region is possible but it
is strongly suppressed as long as $k_B T << e^2/2C_{eff}$.

        The electron transport in an arbitrary single-electron circuit
consisting of tunnel junctions, capacitors, and voltage sources
is described by the ``orthodox'' theory as a sequence of jumps
of single electrons. For any given charge state of the system one should
calculate the tunneling rates for all junctions. In which particular 
junction and at what exactly moment the next tunneling will occur, 
is a matter of chance with the probabilities determined by the 
corresponding rates. After the jump
the charge state changes, and one should calculate all rates anew. These
rates determine the probability distribution for the next jump, and
so on.  This scheme may be used to implement a Monte-Carlo algorithm 
\cite{Bakhvalov} for the simulation of the electron transport.
 Another approach 
\cite{Kulik,Av-Likh-mes} is to solve the kinetic equation
\begin{equation}
\frac{d}{dt} \, p(k)= \sum_{m \neq k} p(m) \, \Gamma (m \rightarrow k)  -
p(k) \sum_{m\neq k} \Gamma (k\rightarrow m) \,  \,\,\, \sum_k p(k)=1 \, ,
\label{kin-eq} \end{equation}
        which describes the evolution of the probability distribution
$p(k)$ among all possible charge configurations.

        The ``orthodox'' theory can also treat the systems with
Ohmic resistances, if they are considerably larger than the 
quantum unit $R_Q$. According to the fluctuation-dissipation theorem
the spectral density of the quantum fluctuations of current through
the Ohmic resistance $R_0$ is $S_I(\omega)=2\hbar\omega /R_0$ (at zero
temperature). Corresponding r.m.s.\ fluctuations of the charge can
be estimated as $\Delta q \sim (S_I(\omega ) \Delta \omega)^{1/2}/
\omega \sim (\hbar /R_0)^{1/2}$ for $\Delta \omega \sim \omega$.
Hence, inequality $R_0 >> R_Q$ allows to neglect the quantum fluctuations
of the charge, $\Delta q << e$. Such an
Ohmic resistor is considered in ``orthodox'' theory as an open circuit
when the effective capacitances are calculated. (One can say that
the charge transferred through $R_0$ ``during'' the tunneling event is
negligible, $(e/C)/R_0 \times \hbar /(e^2/C) <<e$.) 
The charge transfer through the resistor during the time between 
tunneling events leads to
the gradual change in time of the tunneling rates (this change is
stochastic at finite temperature due to Nyquist noise).

        Despite its simplicity, the ``orthodox'' theory of 
single-electronics is sufficient to describe most experimental 
results quantitatively. Among the most important developments beyond
the ``orthodox'' theory  let us mention the account of arbitrary
electrodynamic environment \cite{Nazarov,Nazarov-Ingold} (in 
particular, arbitrary ohmic resistances), the theory of
simultaneous tunneling (cotunneling) in several junctions 
\cite{Aver-Odin,Aver-Naz}, and the account of energy quantization
\cite{Aver-Kor,Kor-dots,Beenakker}.

	\newpage
        \section{Single-electron transistor}

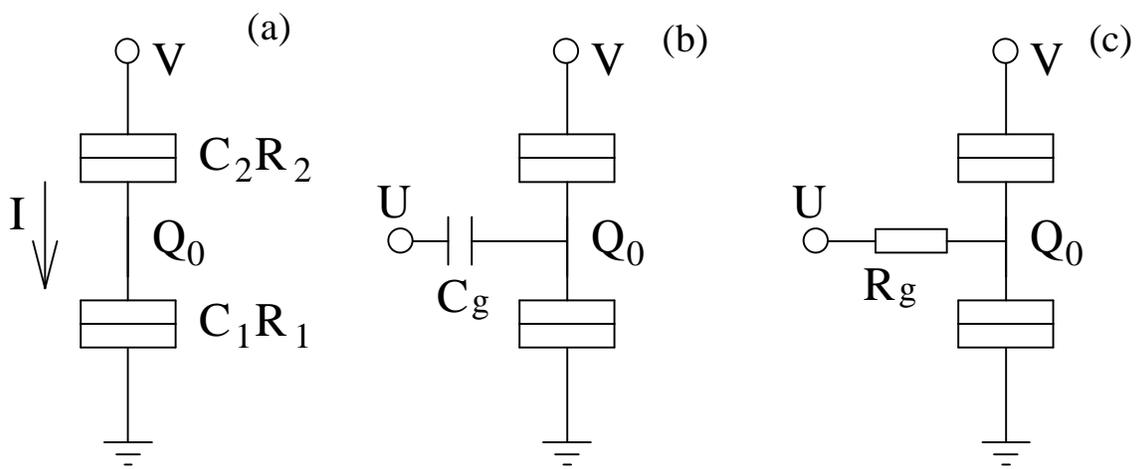
\begin{figure}
\caption{Single-electron transistor (SET): (a) the basic part consisting
of two tunnel junctions in series, (b) capacitively coupled SET (C-SET),
and (c) resistively-coupled SET (R-SET). The current through the SET 
depends on the subelectron fraction of the charge $Q_0$.}
\label{SET}\end{figure}

        The most thoroughly studied single-electron device is 
the so-called Single-Electron Transistor \cite{Av-Likh-86,Likh-87} 
(dubbed as SET-transistor or SET) which is the simplest circuit in
terms of fabrication. Its basic part consists of two tunnel junctions 
in series (Fig.\ \ref{SET}a).
Using this example, let us illustrate the use of  the ``orthodox'' theory. 

        The voltage drops $V_{1,2}(n)$ across the junctions are functions
of the number $n$ of excess electrons on the central island
\begin{equation}
        V_j(n) =V \frac{C_1 C_2}{C_j C_\Sigma} + (-1)^j \frac{Q_0-ne}
{C_\Sigma},
\label{V_j}\end{equation}
        where $V$ is the bias voltage, $C_\Sigma=C_1+C_2$ is the total
capacitance of the central island, and $Q_0$ is its initial 
(background) charge. The energy gain $W_j^\pm (n)$ due to tunneling
($\pm$ denotes two different directions of tunneling) can be calculated as
\begin{equation}
        W_j^\pm (n) = e(\pm V_j(n)-e/2C_\Sigma)
\label{W_j}\end{equation}
        because the effective capacitance is $C_\Sigma$ for any tunneling.
The next step is the calculation of the rates $\Gamma_j^\pm (n)$ using
Eq. (\ref{rate}). The stationary solution of the kinetic equation
(\ref{kin-eq}) for
the probabilities $p(n)$ of different charge states $n$ is as follows
\begin{equation}
        p_{st}(n) \times \left( \Gamma_1^+(n)+\Gamma_2^-(n)\right) 
=p_{st}(n+1) \times \left( \Gamma_1^-(n+1) +\Gamma_2^+(n+1)\right) ,
\,\,\,  \sum p_{st}(n)=1,
\label{stat}\end{equation}
and the average (dc) current $I$ can be calculated as
\begin{equation}
        I=\sum_{n} p_{st}(n) (\Gamma_1^+(n)-\Gamma_1^-(n)).
\label{current}\end{equation}

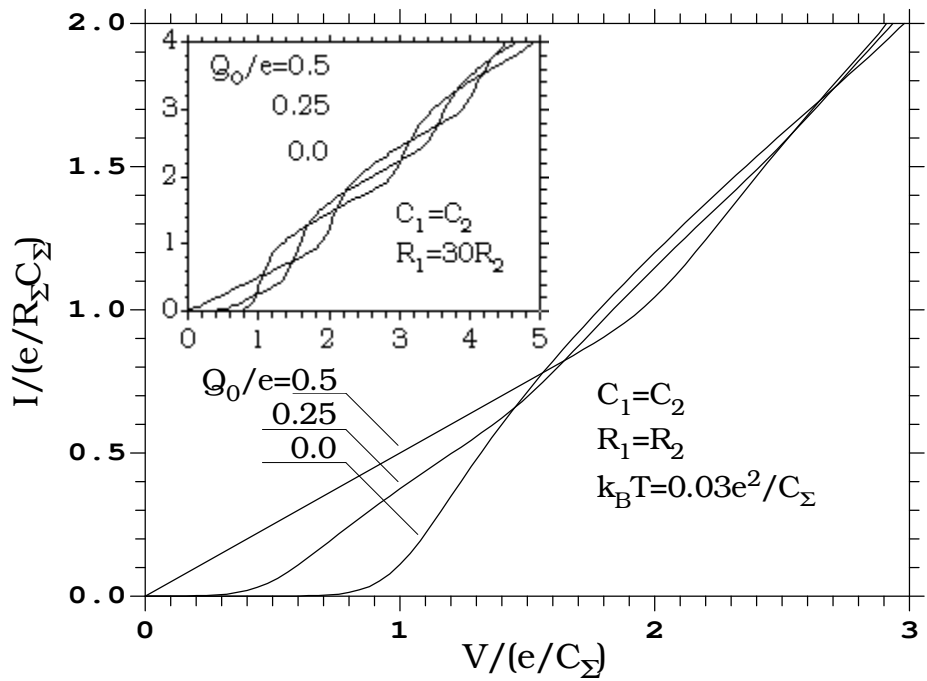
\begin{figure}
\caption{The typical $I-V$ curves for the symmetrical SET and the SET with
different resistances of junctions (inset) calculated using Eqs.\
(\protect\ref{rate})--(\protect\ref{current}) for different $Q_0$.}
\label{SET-I-V}\end{figure}

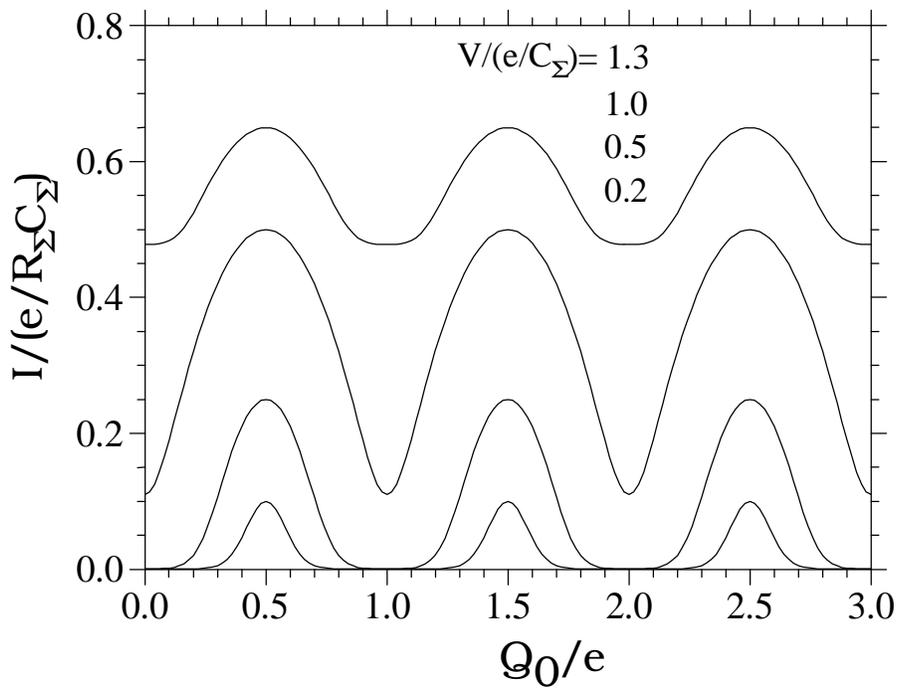
\begin{figure}
\caption{The typical theoretical dependence of the current through 
the symmetrical 
SET  on the induced charge $Q_0$ for different bias voltages $V$.}
\label{SET-I-Q}\end{figure}

        Figure \ref{SET-I-V} shows several dc $I-V$ curves of the symmetric 
double-junction
system ($C_1=C_2$, $R_1=R_2$) calculated in this way. The Coulomb
blockade suppresses the current when the voltage is not sufficient
to provide the energy for single-electron charging of the central island 
(notice also the rounding of the curve cusps due to finite temperature). 
The threshold voltage $V_t$ depends on the background charge,
and its maximal value is $e/C_\Sigma$.
        The same value determines the $I-V$ curve offset at large 
voltages, $I=(V-e/C_{\Sigma})/R_{\Sigma}$.
 The Coulomb blockade
completely disappears ($V_t=0$) for half-integer background charge,
$Q_0=(k+1/2)e$,  because the states with effective
charges $e/2$ and $-e/2$ have equal energies. 
        The current is a periodic function of $Q_0$ 
(Fig.\ \ref{SET-I-Q}) because
the addition of the integer electron charge is compensated by 
the tunneling of one electron in or out of the central island.
This periodic dependence is usually called Coulomb oscillations. Very high
(subelectron) sensitivity to the charge of the central island
is the basis of the SET-transistor
operation. Controlling $Q_0$ by capacitively coupled gate
(C-SET, Fig.\ \ref{SET}b) or via coupling resistor (R-SET, Fig.\ 
\ref{SET}c), one
controls the flow of electrons tunneling through the SET-transistor.

        R-SET is quite difficult to implement because the coupling
resistance $R_g$ should be much larger than $R_Q$ to prevent quantum
fluctuations of $Q_0$; simultaneously the resistor size should be small
so that its stray capacitance does not 
significantly increase $C_\Sigma$. Experimental demonstration of the 
R-SET is still an unsolved problem despite significant progress
in this direction \cite{Kuzmin-Bloch,Pashkin}.
	Similar difficulty does not allow so far the experimental study
of the so-called RC-SET \cite{RC-SET} which would be very useful
in digital circuits because of its multistable characteristics.

        In contrast, C-SET was demonstrated repeatedly by many
scientific groups
using different materials and technologies (see Section II). 
In some laboratories it is  a routine
device which is used to measure very small charge variations, for
example, in other single-electron circuits.
The gate voltage $U$ (see Fig.\ \ref{SET}b) induces the effective charge
into central island, $Q_0 \rightarrow Q_0+C_g U$ ($C_g$ is the
gate capacitance), hence, Fig.\ \ref{SET-I-Q} can be considered as a 
control curve of the C-SET. The gate voltage period is equal to
$\Delta U=e/C_g$. If $C_g$ is comparable to the junction capacitance,
its contribution to the total capacitance should be also taken into
account. To calculate characteristics of C-SET it is sufficient to
use Eqs.\ (\ref{V_j})--(\ref{current}) with the substitution
$C_1 \rightarrow C_1+\alpha C_g$, $C_2 \rightarrow C_2+(1-\alpha)
C_g$, $Q_0 \rightarrow Q_0+C_g U -\alpha C_gV$, where $\alpha$ is
an arbitrary number (usually $\alpha=0$ or $\alpha=1$ is used).

        If the resistances of two junctions of the C-SET are 
considerably different, then the $I-V$ curve shows substantial 
periodic oscillations with period $\Delta V =e/C_1$ 
(for $R_1 \gg R_2$), called the Coulomb staircase.
The cusps of the $I-V$ curves shown in the inset of Fig.\ \ref{SET-I-V} 
correspond 
to condition $W_2=0$ in Eq.\ (\ref{W_j}). Each period of staircase 
corresponds to an additional electron on the central island.
Coulomb staircase is typical for experiments with the use of scanning
tunneling microscope because the tunnel junction between its tip 
and conducting particle is typically much smaller than the 
junction between particle and substrate.
In contrast, Coulomb staircase in C-SETs made of metal
films is usually very weak, because this technology is able to 
produce junctions of the same parameters
(in this case the staircase may be the evidence of a bad sample).

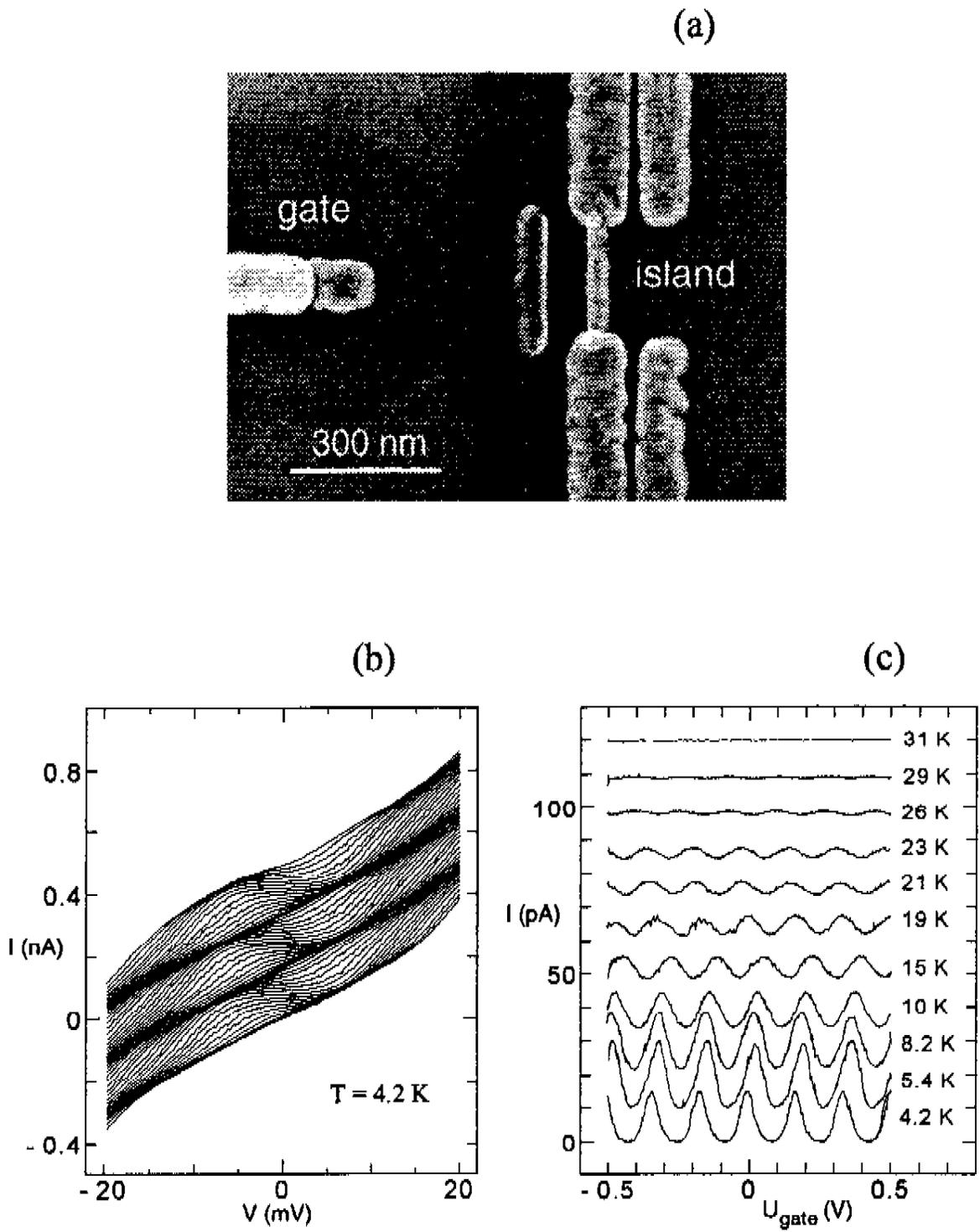
\begin{figure}
\caption{Experimental realization \protect\cite{Tsai} of the C-SET 
using narrow metal films: (a) layout, (b) $I-V$ curves for different
gate voltages and (c) the dependence of the current on the gate 
voltage at different temperatures (courtesy of Y. Nakamura). 
The curves in (b) and (c) are 
shifted vertically for clarity.  }
\label{Tsai-fig}\end{figure}

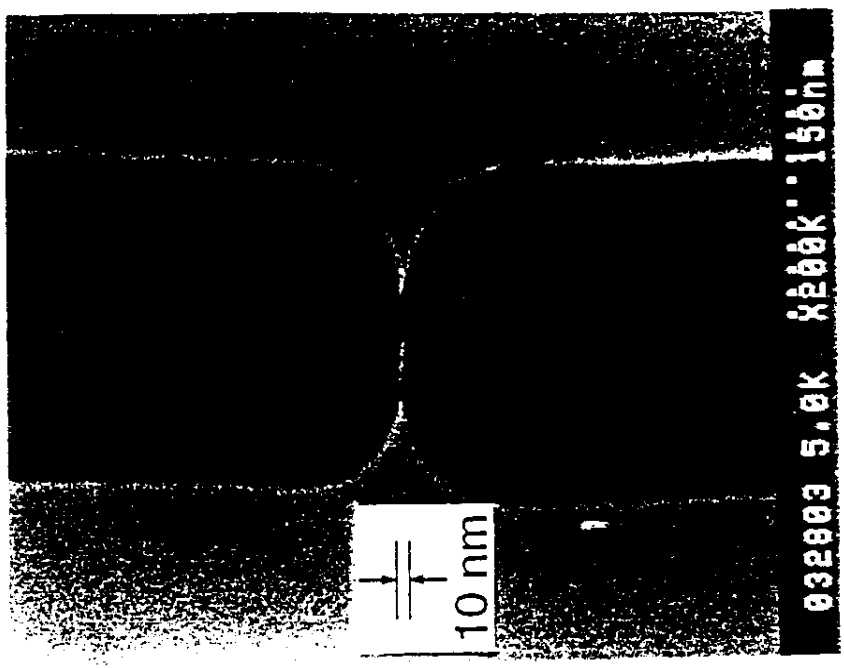
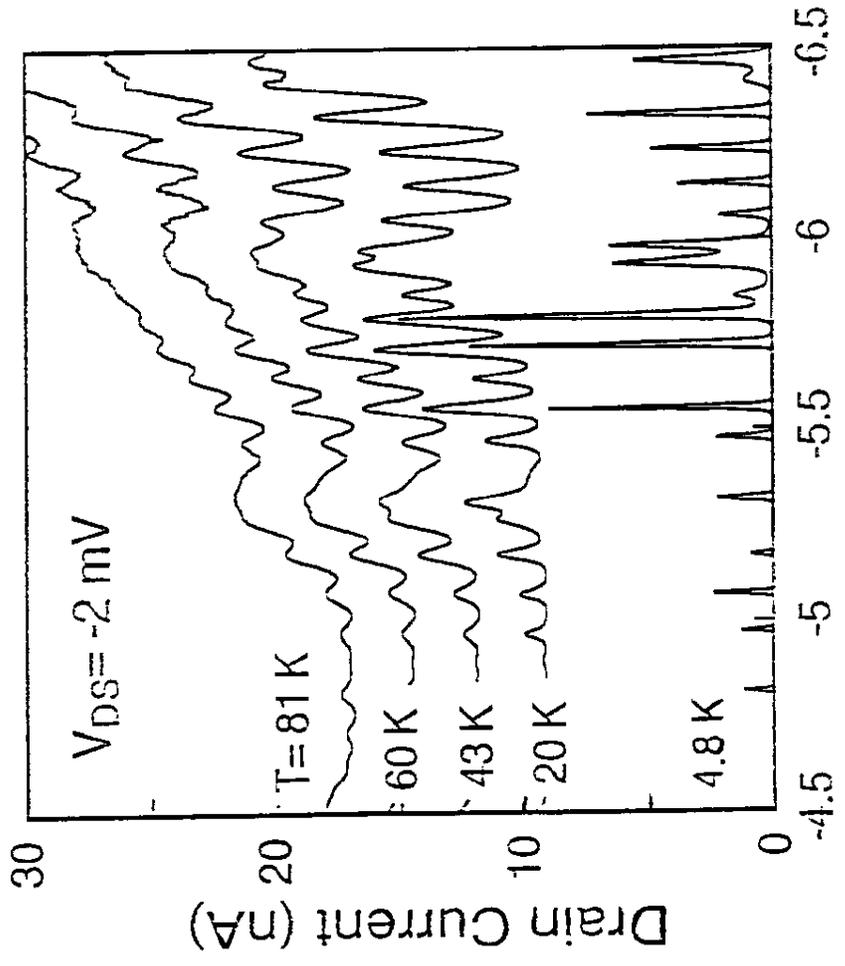
\begin{figure}
\caption{Si-based SET-transistor \protect\cite{Chou}: electron micrograph of
the structure and the dependence of the current on the gate voltage
(courtesy of E. Leobandung). The current is actually due to the hole
tunneling, so the structure is named Single Hole Quantum Dot Transistor.}
\label{Chou-fig}\end{figure}

        The theory of the SET-transistor is very well confirmed 
experimentally. Figure \ref{Tsai-fig} shows the example of the
layout, $I-V$ curve, and the dependence of the current on the gate 
voltage for the
C-SET made of metal films \cite{Tsai}. Unusually high operation
temperature (up to 30 K) was achieved in this experiment by the use of 
the film anodization; typically metal film SETs operate at $T<4$K
so far. Simple ``orthodox'' theory described above is usually
sufficient for the good quantitative agreement with experimental data
for metallic C-SETs. Some additional factors should be typically taken
into account for semiconductor C-SETs and double-junction structures 
with a molecule as the central island (see Section VI). Figure 
\ref{Chou-fig} shows the SEM image and the current -- gate voltage
dependence for the recently demonstrated Si-based SET \cite{Chou}
operable at the temperature over 80 K.
	The current in this device
was actually carried by the tunneling holes, and the energy level 
spacing was comparable to the Coulomb blockade energy, that is why 
the authors of Ref.\ \cite{Chou} call this device Single Hole Quantum 
Dot Transistor. Notice that the Coulomb
oscillations in Fig.\ \ref{Chou-fig} are not exactly periodic and they
are superimposed on the monotonic dependence
on the gate voltage. We will discuss the reason for this difference
from the behavior of metallic SETs in Section VI.

	The C-SET can be used as a highly sensitive electrometer.
        The charge sensitivity of the SET-transistor is limited by 
its noise. In
experiments the spectral density of this noise has usually $1/f$
dependence \cite{Martinis-noise,Verbrugh,Visscher,Pashkin-Cr} 
that can be explained as the random capture
of electrons by impurities in the tunnel barriers and/or the 
substrate in the vicinity of SET-transistor. The noise
decreases with the improvement of the technology. The lower limit is 
determined by the intrinsic thermal/shot noise of the SET-transistor
\cite{Kor-noise,Hershfield} which was recently measured \cite{Birk}
at relatively high frequency where the contribution of $1/f$ noise
is small. The ultimate sensitivity \cite{Kor-noise} limited
by the intrinsic noise of the C-SET is given by $\delta Q_0 \approx
2.7 e (k_BTC_\Sigma/e^2)^{1/2}(RC_\Sigma\Delta f)^{1/2}$ where $\Delta f$
if the bandwidth. For typical parameters of the present day experiments,
$C_\Sigma \sim 3 \times 10^{-16} F$, $T\sim 0.1 $K, $R \sim 10^5 \Omega$,
the ultimate sensitivity is about $2\times 10^{-6} e/\sqrt{Hz}$, while
the best experimental sensitivity recorded so far is $7\times 10^{-5} e/
\sqrt{\mbox{Hz}}$ at 10 Hz \cite{Visscher}.

        Fitting of the experimental results obtained at $T<0.1$ K usually
shows that the real temperature is larger than the temperature of the
cryostat. This can be explained as the heating due to the transport current
\cite{Kautz-heat,Kor-heat,Verbrugh} and imperfect microwave isolation
of the SET-transistor from room-temperature environment.

        The ``orthodox'' theory should be modified for
the calculation of the small current well below the Coulomb blockade
threshold at low temperatures. In this case the single electron tunneling
is blocked, and the current is due to ``simultaneous'' tunneling 
(cotunneling) of two electrons through both junctions 
\cite{Aver-Odin,Aver-Naz}. Because of the quantum
nature of the process involving the whole electrical circuit, cotunneling
is also called Macroscopic Quantum Tunneling of charge
($q$-MQT). The rate of such a 
process is proportional to the product $(R_Q/R_1)(R_Q/R_2)$, and, hence,
is relatively small for $R_i \gg R_Q$.

	\newpage
        \section[digital]{Digital single-electron devices}

        The most important potential application of the single-electronics
is the integrated digital electronics which could substitute 
conventional semiconductor transistor technology at the size scale 
below 30 nm. There
have been many theoretical suggestions on this subject, we will discuss
several of them.

        \subsection[SET]{ Logic/memory using SET transistors}

        Conceptually the simplest way to realize digital 
single-electronics is to use SET-transistors instead of FET transistors
in circuits resembling conventional electronics. It is possible
to use capacitively coupled (C-SET) or resistively coupled (R-SET)
transistors. Because R-SET is still too difficult for fabrication,
let us limit the discussion to C-SET circuits.

        For C-SET the dc input current is zero, hence, the power
amplification is formally infinite. However, the voltage gain
$K_V$ is not large \cite{Likh-87} 
(in contrast to semiconductor MOSFET transistors),
\begin{equation}
        K_V \le C_g/\min (C_1, C_2).
\end{equation}
        The condition $K_V>1$ which is necessary for the operation 
of logic devices requires the gate capacitance $C_g$ to be larger than 
the junction capacitance.

        The buffer/inverter can be realized by one SET-transistor
in series with the load resistor $R_L$. Notice that the fabrication
of such a resistor is not a big problem in contrast to the resistor for
R-SET because there is no limitation on its stray capacitance.
However, to reduce the number of technological steps, it is more
reasonable
to use a tunnel junction instead of the load resistor \cite{loadjunction}.
        Calculations show \cite{Kor-trans} that for good operation
of buffer/inverter, $R_L$ should be at least 10 times larger than the
junction resistance. Hence, the additional power dissipation in the
load resistor will be much larger than that in SET-transistor.

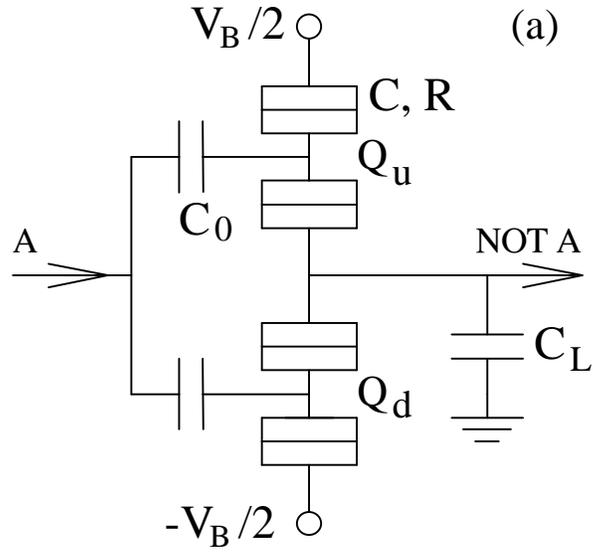
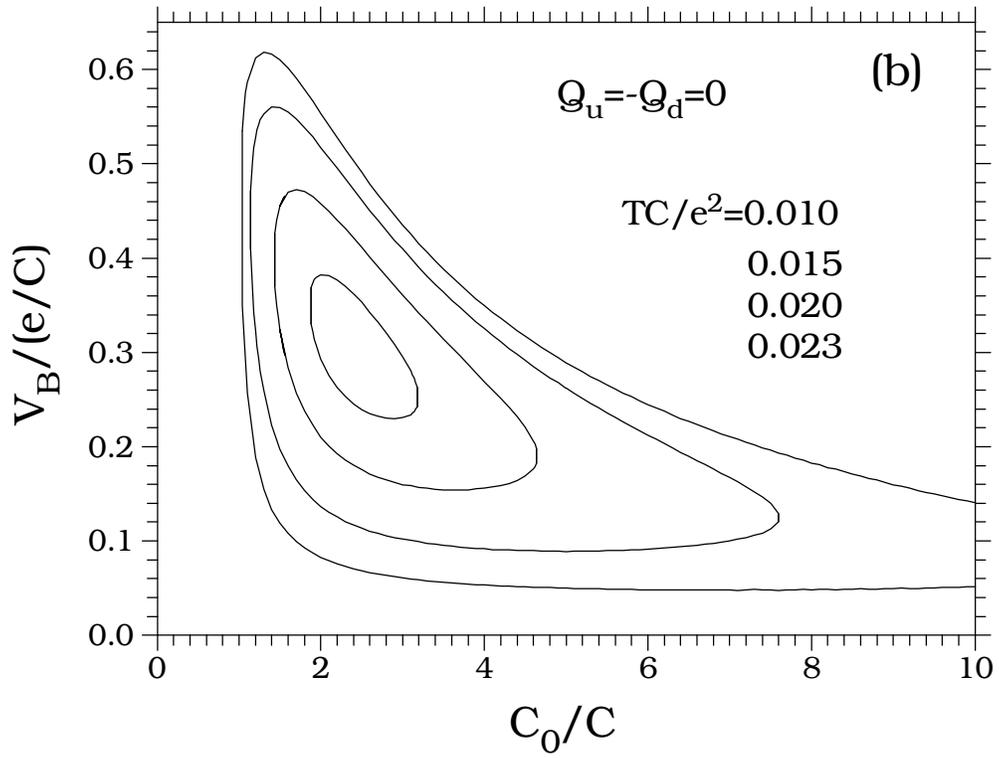
\begin{figure}
\caption{(a) The complementary inverter made of two SET-transistors 
and (b) its parameter window for different temperatures 
\protect\cite{Kor-trans}.}
\label{complem}\end{figure}

        To reduce the power consumption the complementary circuits can 
be used \cite{Likh-87,Tucker,Kor-trans,Chen}. It is important that in contrast
to CMOS technology in which n-MOS and p-MOS transistors are physically
different, both complementary SET-transistors can be physically
identical 
because of the periodic dependence of the current on the gate voltage. 
To achieve the
complementary action, the operating point of one transistor should be on 
the raising branch of this dependence while for the other transistor
it should be on the falling branch. It can be done with the use
of additional capacitors \cite{Tucker} or different background charges
in complementary transistors \cite{Kor-trans} (Fig.\ \ref{complem}a). 
However, even without any special effort,
complementary action occurs automatically in the simplest case of two
symmetrical transistors with zero background charges \cite{Kor-trans}. 
It is interesting that in terms
of the maximal operation temperature this simplest case is very close
to the optimal one.

         The maximal temperature at which the complementary inverter 
still amplifies the signal is equal to $0.026 e^2/Ck_B$ 
\cite{Kor-trans} where $C$ is the capacitance of one tunnel
junction. Notice that the same maximal temperature can be obtained
for resistively loaded transistor if $R_L$ is very large. The maximal
temperature is achieved when the gate capacitance $C_g$ is about 
twice larger than the junction capacitance. The optimal $C_g$
(corresponding to largest parameter margins)
increases when the temperature decreases, so that $C_g/C \approx 3$ 
seems to be more or less the  best  choice  for  experimental  
realization.  To  have reasonable
parameter margins, the temperature should be crudely twice less than
the maximal temperature, $T \sim 0.01 e^2/Ck_B$. In this case the 
margins for bias voltage and $C_g$ are sufficiently wide (see Fig.\
\ref{complem}b), and the critical margin is that for the fluctuations 
of background charges (about 0.1 $e$).

        The operation point which optimizes the maximal temperature
of the complementary inverter corresponds to relatively large
power consumption about $2 \times 10^{-3} e^2/RC^2$
per SET-transistor. However, in the ``power-saving'' mode
for a price of slight reduction of the operation temperature, the
power consumption can be reduced down to $10^{-4} e^2/RC^2$ per
transistor \cite{Kor-trans}.

        The switching time of the complementary inverter is
close to $3RC_L$ where $C_L$ is the load capacitance (see Fig.\
\ref{complem}a). Relatively large load capacitance, $C_L \agt 300 C$, should
be used in order to make negligible the fluctuations of the output
voltage due to the shot noise in the transistors.

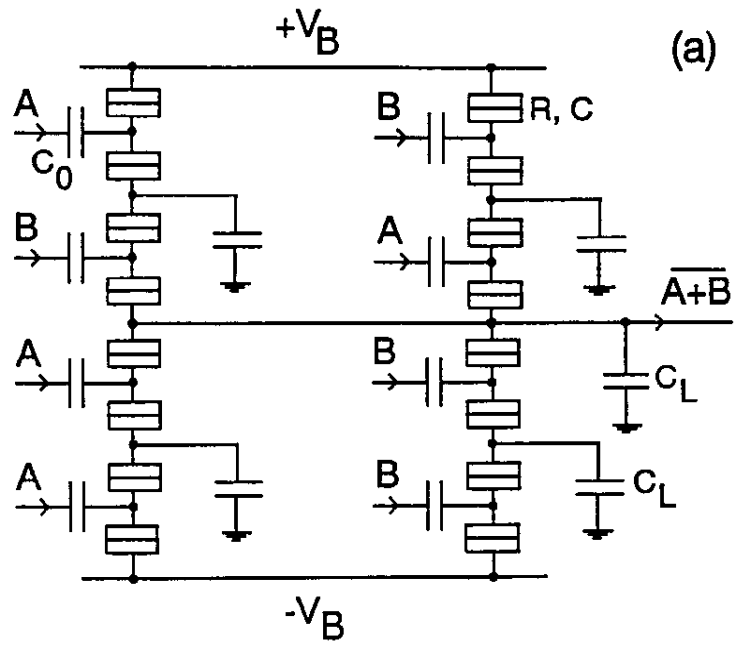
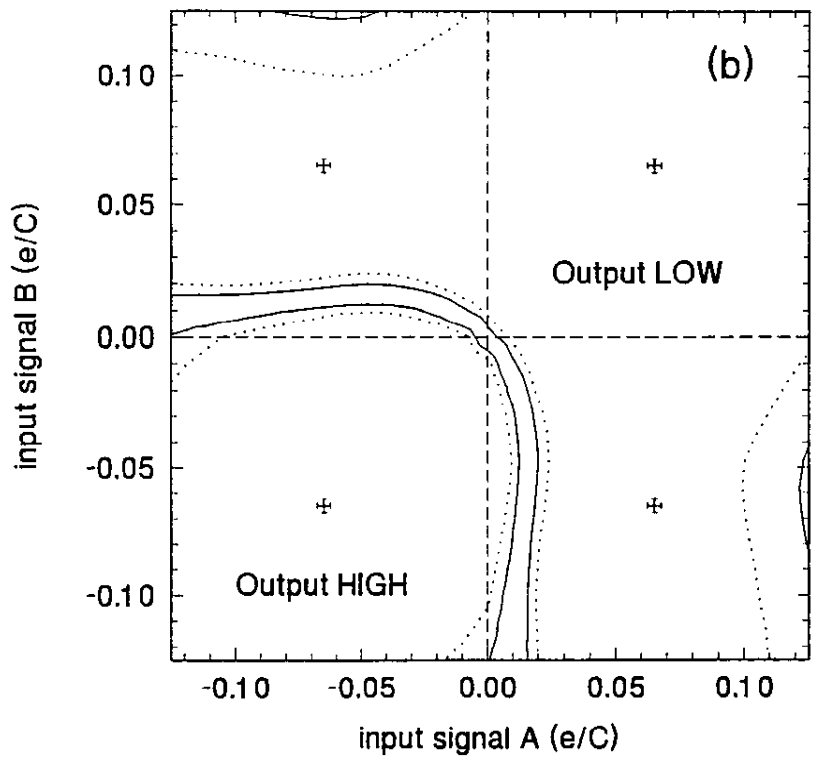
\begin{figure}
\caption{ (a) The NOR gate made of SET-transistors and (b) its typical
output characteristics on the plane of input signal amplitudes
\protect\cite{Chen}. The solid lines in (b) show the ``active'' region
where the output cannot be definitely interpreted, and the areas between
solid and dashed lines correspond to the noise margins.}
\label{SET-NOR}\end{figure}

        Two inverters connected in a ``circle'' constitute the bistable
flip-flop which can be used as a static memory cell. Almost all results 
of analysis of the inverter are directly applicable to the flip-flop. 
Slightly lower temperature and slightly narrower parameter
margins are required for the operation
of the logic gates based on SET-transistors \cite{Chen}.
A possible structure of the NOR gate \cite{Chen} is shown in Fig.\ 
\ref{SET-NOR}a. 
Notice that in contrast to SET inverter which is similar to the circuit
used in conventional digital electronics, design of the SET NOR gate
differs from the conventional one. (The direct reproduction of the design 
is impossible because of different characteristics of SET and MOSFET 
transistors.) The operation of SET NOR gate is illustrated in Fig.\ 
\ref{SET-NOR}b.
One can see that the threshold lines are close to the perfect (square)
shape. Inversion of the bias voltage transforms NOR gate into NAND
gate with the similar characteristics. NOR and NAND gates accompanied 
by the NOT gate (inverter) are more than sufficient for performing
arbitrary logic functions. However, special design for some other 
gates, for example, SET XOR gate \cite{Chen} can help to make the
logic more efficient. The single-electron transformer {Av-Kor-Naz}
can also be useful in the SET-transistor logic.

        For a technology with a minimal feature size of 2 nm
one can expect the capacitances of the tunnel junctions as low
as $3 \times 10^{-19}$ F. This corresponds to $e^2/Ck_B=6\times
10^3$ K; hence, the maximal temperature at which SET-transistor
still amplifies the signal is close to 150 K. It would allow
the operation of the SET logic at the liquid nitrogen temperatures.
(We see that the room temperature operation requires the fabrication
technology at sub--1 nm level.)
For the estimate of the typical switching time let us take $R \approx
300 k\Omega$ and $C_L\approx 10^3 C\approx 3\times 10^{-16}$ F,
then this time is about 1 ns. The power consumption per transistor
is quite small, about $2\times 10^{-8}$ W for the parameters above 
in a typical operation point (in a ``power saving operation point'' it 
is about $10^{-9}$ W). However, because the density is very large,
the power dissipation is a serious problem. For example, at
$10^{11}$ transistors per cm$^2$ even in the power saving mode the
total power is on the order of 100 W/cm$^2$. Probably, even more difficult
problem of the logic/memory based on SET-transistors is the necessity
to keep fluctuations of background charge within the margins on the
order of $0.1 e$. This is a common problem for any integrated 
single-electronics, we will discuss the possible solutions in a 
separate subsection.

        Let us emphasize that a single SET logic device can be 
relatively easy fabricated using the present-day technology. The
multilayer technology which allows relatively large gate capacitances
and solves the problem of connections between circuit elements has been
already developed \cite{Martinis-trans,Visscher}. One can expect
that first SET logic devices will be demonstrated within a few
years (of course they will probably operate at $T<1$ K and require 
individual adjustment of background charge at each island).

        \subsection[SEL]{SEL logic and single-electron trap}

        In the logic/memory based on SET-transistors the logical
unity and zero are represented by different dc voltage levels similar
to how it is done in conventional digital electronics. Another
possibility is to represent bits by single electrons
\cite{Likh-Sem,Likh-Polon,Av-Likh-log,Naz-Vysh,Nakazato-log,Ancona},
so that one extra electron in a conducting island would correspond
to logical unity, while the absence of an extra electron, to 
logical zero. The circuits based on this truly single-electron approach
are called Single-Electron Logic (SEL) \cite{Likh-Sem,Likh-Polon,Av-Likh-log}.
The apparent advantage of this idea is a low power
dissipation because in a static state there is no current, and the 
logical processing of one bit of information requires only few tunneling
events.

        In the initially proposed SEL logic \cite{Likh-Sem} single 
electrons propagate together with information along the relatively
long arrays of tunnel junctions and ohmic resistors. In this scheme 
the proper dc biasing is
a difficult problem because the bias should be distributed in a
specific way among the large number of cells. To resolve this problem
it was suggested \cite{Likh-Polon} to separate the propagation 
of electrons and information:
electrons tunnel across the elementary cell which is a short biased array of 
junctions, while the information propagates from one cell to another 
perpendicular to the motion of electrons.

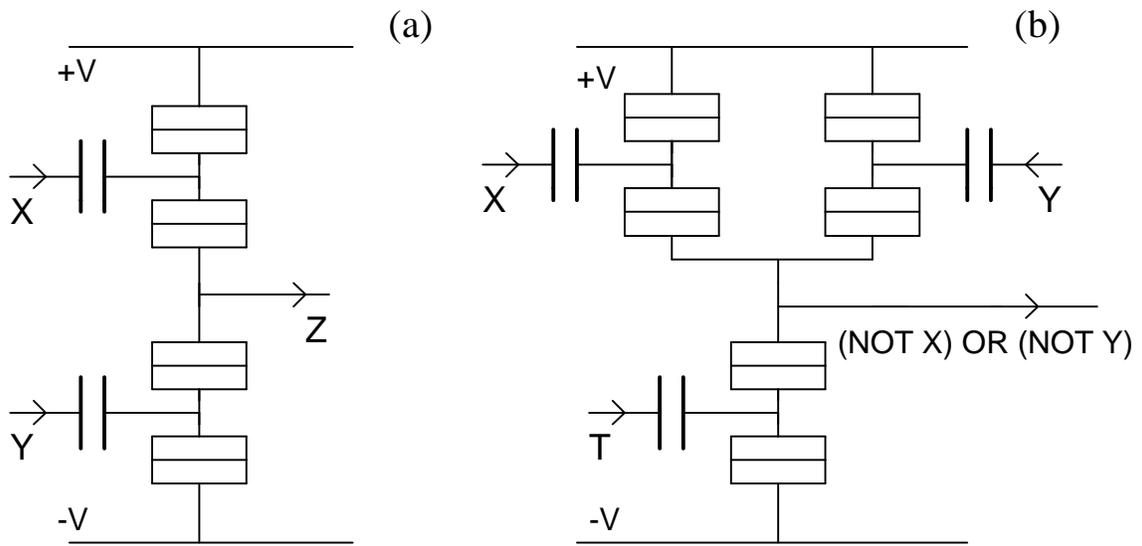
\begin{figure}
\caption{The basic cell of the SEL logic and (b) the SEL NOR gate 
\protect\cite{Likh-Polon,Av-Likh-log,Naz-Vysh}.}
\label{SEL}\end{figure}

        Fig.\ \ref{SEL}a shows the basic cell of the SEL family 
considered in Refs.\
\onlinecite{Likh-Polon,Av-Likh-log,Naz-Vysh}. Notice that it is 
similar to the complementary SET inverter, however, the important
difference is that the capacitance of the middle island of a SEL cell
is on the order of the junction capacitance (in contrast to large $C_L$
in SET inverter). Inputs X and Y determine the charge state of the middle
island. For example, if the lower branch of the cell is ``closed'' by the
signal Y, and the signal X opens the upper branch of the cell, then one
extra electron tunnels through the upper branch to the middle island.
This creates the logical ``unity''. Parameters are chosen in a way
that the next electron cannot come because of the increased potential
of the island. The extra electron can be removed (creating logical ``zero'')
from the middle island by closing the upper branch and opening the lower 
one. The charge of the middle electrode being the output of the cell,
is used to affect the charge state of the next cell.

        Figure \ref{SEL}b shows the SEL logical gate NOR. Signals X and Y
are logical inputs. The middle island becomes charged by an extra 
electron when one of the upper branches opens. Clock signal T discharges
the middle island at the end of the clock signal (upper branches should
be closed at this time).

        In contrast to the SET-transistor circuits, there is a strong
back action from the output to the input in SEL circuits. Numerical
simulations has proved \cite{Likh-Polon,Av-Likh-log,Naz-Vysh} that
the proper choice of parameters provides the unidirectionality of
the signal propagation. However, because of the back action,
the parameter margins are considerably narrower than in 
SET-logic case. 

        Another problem of SEL logic is that the information
coded by a single electron can be destroyed by a single erroneous
event due to cotunneling or thermoactivated tunneling. 
The possible solution would be the use of multijunction arrays as 
branches of SEL circuits, however, this possibility was not studied
quantitatively yet.

        The problems mentioned above make SEL logic circuits much more
difficult to implement than SET-transistor logic at least at the 
present stage. However, one can hope that the problems will be 
eventually solved by the search of the optimal design and the improvement 
of the technology. This hope is strongly supported by a successful
experimental demonstration \cite{Fulton-trap,LaFarge-trap,Likh-trap}
of the ``memory cell'' in which logical bit is represented by a single
electron on the conducting island.

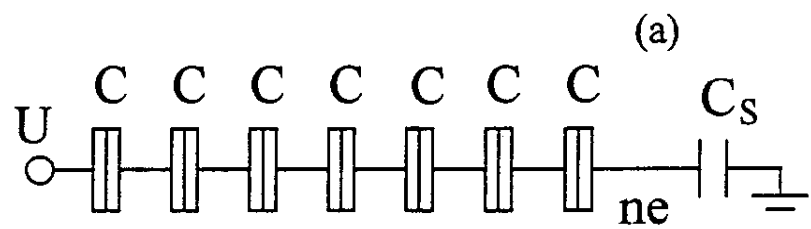
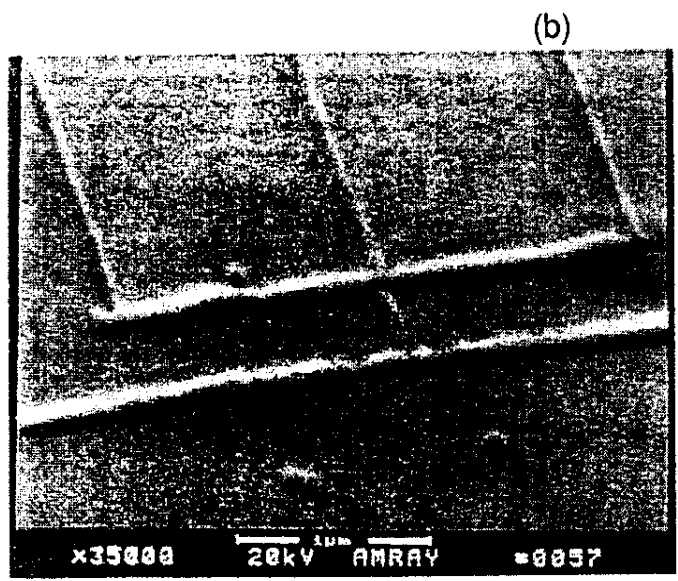
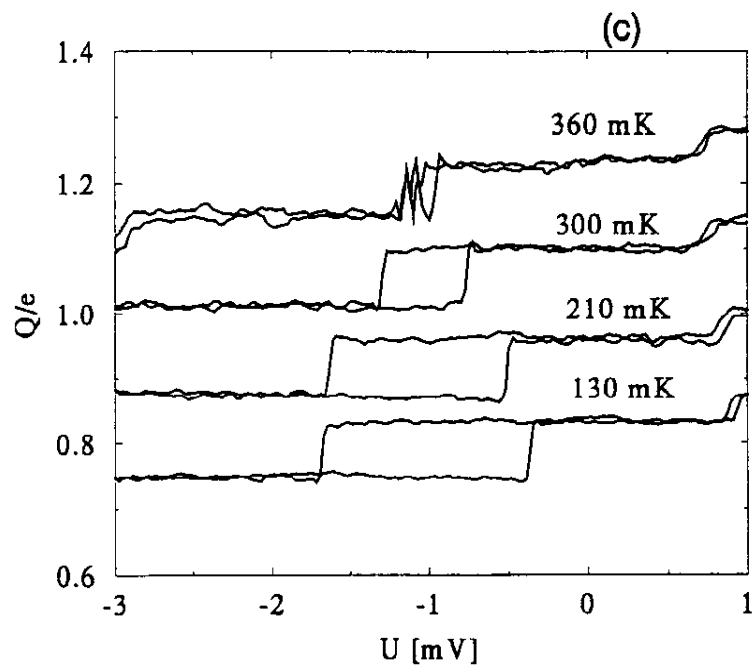
\begin{figure}
\caption{The single-electron trap: (a) schematic drawing,  (b)  the 
electron
micrograph of the structure and (c) the hysteretic dependence of the trapped
charge (multiplied by the coupling coefficient) on the voltage $U=V_{trap}$ 
\protect\cite{Likh-trap} (courtesy of P.\ D.\ Dresselhaus). The charge is 
measured by the SET-transistor 
(upper part of the layout). The height of each loop in (c) corresponds 
to one extra electron in the trap.}
\label{trap}\end{figure}

        This circuit which is usually called ``single-electron trap'',
consists of several (typically, 5--7) tunnel junctions in series with 
a capacitor (Fig.\ \ref{trap}a). Experimentally (Fig.\ \ref{trap}b), 
this is an array of metal junctions which
ends with a relatively large island so that its capacitance to the ground 
$C_S$ is  comparable  to  the  junction  capacitance $C$. 
The number of electrons on the island can be changed by application of
the bias voltage $U$ (Fig.\ \ref{trap}a).  Several
charge states can be stable for the same $U$ because of 
the Coulomb barrier created by the array of junctions.
In the case of zero background charges the tunneling is blocked when
$|V|<V_t=Ne/2C_{eff}$, $C_{eff}=C+((N-1)/C+1/C_S)^{-1}$ where $V$ is the
voltage across the array consisting of $N$ junctions.
 One additional electron in the edge island changes $V$ 
by $\Delta V=e/(C_S+C/N)$. Hence, as many as $m=1+\mbox{int}(2V_t/\Delta V)$ 
different states can be within the Coulomb blockade range. 
Two stable states ($m=2$) which differ by one electron on the edge
island represent logical unity and zero in a single-electron trap (Fig.\
\ref{trap}c).

        Similar to the SEL logic circuits, the erroneous switching of the 
single-electron trap are due to thermoactivated processes and cotunneling.
Both processes are suppressed with the increase of the number of junctions
in the array. The error rate less than 1 switching per 12 hours was 
demonstrated at the temperature of 50 mK in the 7-junction array made of 
aluminum tunnel junctions \cite{Likh-trap}. The charge state of the island 
was monitored with a help of near-by single-electron transistor (Fig.\
\ref{trap}b). Theoretical consideration shows \cite{Fonseca} that 
in principle the error rate below $10^{-17}$ s$^{-1}$ can be achieved
in a similar trap.

        If the capacitance $C_S$ of the storage island is relatively large 
so that $e/C_S$ is considerably smaller than the Coulomb blockade threshold 
$V_t$, then there are many, $m\approx 2V_t/(e/C_S)$, stable states within 
the blockade range. 
Representation of the logical bit by a single electron is ineffective
in this case, however, the bit can be stored as several electrons on the
island. For example, $q=+me/2$ can correspond to unity, and $q=-me/2$
corresponds to zero. The power dissipation during the writing process
is larger than in single-electron case, however, for $m \sim$10--100 
it is still extremely small.
	The advantage of the multi-electron storage is that 
single erroneous tunneling events do not destroy the information, and
hence, the simple refreshing of information can be used to avoid errors
(in the single-electron case refreshing is possible only with the use of
redundancy).

        The multi-electron storage based on the Coulomb blockade was
demonstrated \cite{Nakazato-mem} using a side-gated constriction in
$\delta$-doped layer of GaAs. The arrays of tunnel junctions appeared 
naturally in the constriction due to disorder. Several tens of electrons 
were used to represent a bit. The operation was confirmed up to the
liquid helium temperature (4.2 K), and the storage time was as long as 
several hours.

    The single-electron memory effects at room temperature were reported
in silicon-based structures \cite{Yano}. The current through the narrow
ultrathin poli-Si film showed the hysteresis as a function of gate voltage.
The effect was ascribed to the trapping of single electrons in small 
naturally formed grains of poli-Si. The use of disorder for the
creation of extremely small islands (far beyond the limits of the modern
lithography) offers the possibility of the high temperature operation.
However, such a technique obviously has a problem with the reproducibility
of sample characteristics because of the random nature of the island
creation.

        \subsection[wireless]{Wireless Single Electron Logic}

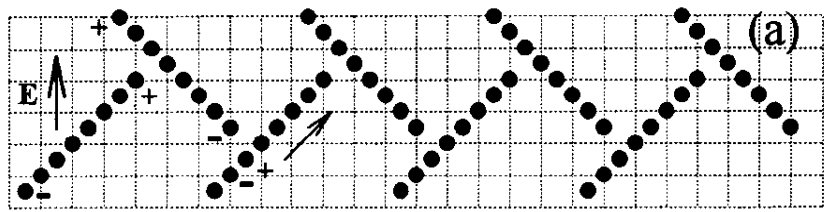
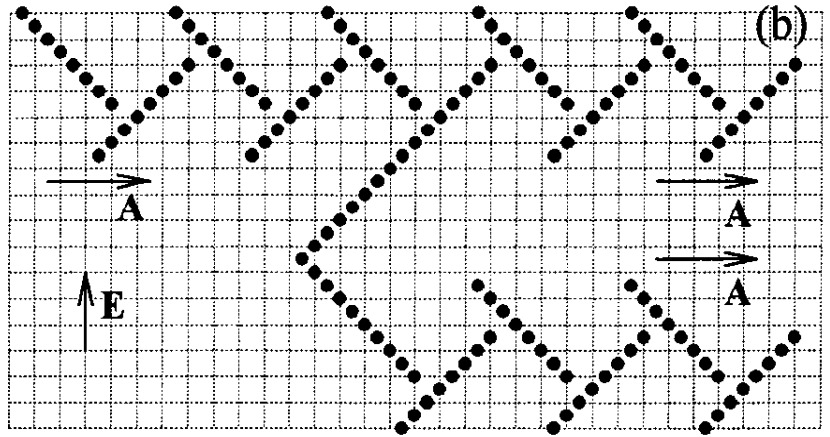
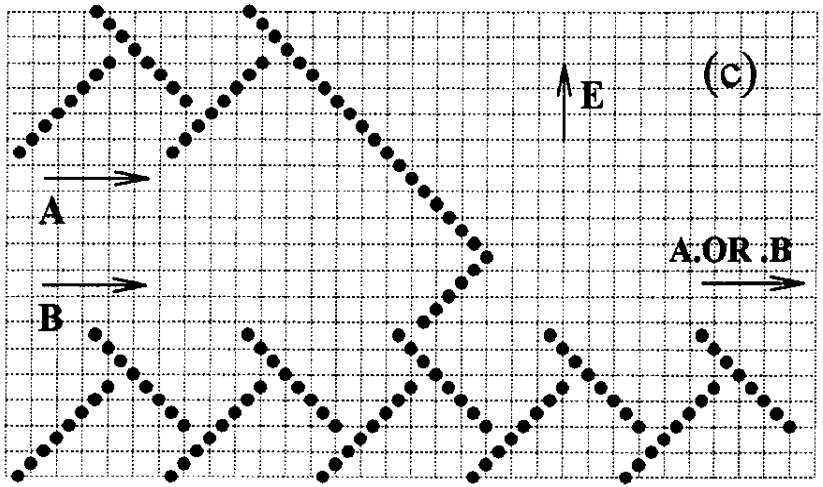
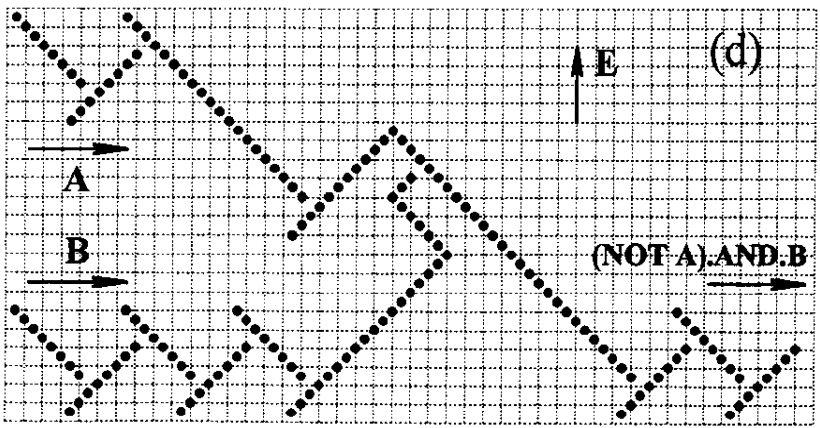
\begin{figure}
\caption{``Wireless'' Single-Electron Logic \protect\cite{Kor-isl} based on
tunneling between small conducting islands and biased by electric field
$E$. (a) The propagation line, (b) the circuit for fan-out, (c) the
logical gate OR (gate AND has a similar design), and (d) the gate
(NOT A).AND.B which can be used as an inverter.}
\label{WISE}\end{figure}

        Both SET-transistor circuits and SEL logic considered above
require wires for the power supply and connections between circuit elements.
Though the necessity of wires is not a principal problem, it is
obviously inconvenient at the few-nanometer size scale. In the Wireless
Single Electron Logic (sometimes the abbreviation WISE is used) proposed 
in Ref.\ \onlinecite{Kor-isl}
the power is supplied by alternating external electric field,
and the capacitive coupling between neighboring cells is due
to their close location.
        The ``device'' consists of many conducting islands, and the
logical functions are determined by their specific arrangement (Fig.\
\ref{WISE}).  Small ``puddles'' of 2D electron gas, small metallic droplets 
on an insulating substrate,
or conducting clusters in a dielectric matrix are possible
implementations of the islands.
        The basic cell of the logic is a short chain of closely
located islands so that electrons can tunnel between neighboring
islands. There is no tunneling between different chains because of
the larger separation.
        
        Application of in-plane electric field $E$ creates the voltage
between the islands. When $E$ exceeds the Coulomb blockade
threshold $E_t$, the tunneling occurs somewhere inside the chain,
producing an electron-hole pair. The electric field drags the 
components of the pair apart
towards the opposite edges of the chain, creating the polarized state. 
        If now the field $E$ is decreased, the pair eventually
annihilate, however, it will occur at the field $E_a$ considerably smaller
than $E_t$. Stability of both polarized and nonpolarized states
for $E$ between $E_a$ and $E_t$ allows to use these states as logical
zero and unity.

        The polarization change can propagate along a line of closely 
located
chains (Fig.\ \ref{WISE}a). Suppose that all chains are not polarized
initially, and $E$ is slightly less than $E_t$. This is a metastable state.
If one chain becomes polarized, the field of extra electron
(hole) on the edge island increases the potential difference between 
neighboring islands of the next chain (Fig.\ \ref{WISE}a). 
This makes tunneling
energetically favorable and leads to polarization of the next chain.
This in turn polarizes the next chain and so on.
        The unidirectional propagation (in Fig.\ \ref{WISE} from 
left to right) is a consequence of the asymmetry of the circuit.

        The natural fan-out of the signal into two lines can be 
realized if both edge islands of a chain
are used to trigger the next chains (Fig.\ \ref{WISE}b).

        A ``bi-controlled'' chain (fifth from the right in Fig.\ 
\ref{WISE}c) which can be triggered by the polarization of 
either of two neighboring input chains can be used as the basic part 
of the logical gate OR.
        The logical gate AND can be designed similar to the OR gate,
but with slightly larger distance between the ``bi-controlled'' chain 
and the neighboring input chains, in order to decrease their influence.
Another possibility is to 
make the islands of ``bi-controlled'' chain slightly smaller in order 
to increase the Coulomb blockade energy.

        Because of the asymmetry between logical zero and unity the
design of the inverter is relatively complex.
The circuit shown in Fig.\ \ref{WISE}d
implements the logical function (NOT A).AND.B if the signal 
from input A comes before the signal from input B.
        This circuit can be used as NOT A, if logical unity always 
comes from input B and it comes later than signal A.

        According to numerical simulations,
the correct operation of the circuits shown in Fig.\ \ref{WISE} requires 
that the magnitude of
external field $E$ lies within $5\%$ margin \cite{Kor-isl}. This number
gives also a crude estimate of the margins for other parameters 
(fluctuations of radius, spacing, etc.)

        Considered logical gates together with propagation lines 
and fan-out circuits, are sufficient for computing.
        In the simplest mode of operation, all chains inside a device 
initially have zero polarization and external field is zero.
Then external field increases up to a value for which all gates 
operate correctly,
and cells start to switch in accordance with the input information
flowing from the edges of the device. The result of the computation is
the final polarization of output cells which can be read out, for 
example, by single-electron transistors.
        This simplest mode of operation can obviously be improved
by the use of periodic changes of the external field (``clock cycles'').
Properly chosen levels of the field can reset some cells but
preserve the information in other cells.

        Let us notice that the Wireless Single Electron Logic proposed
in Ref.\ \onlinecite{Kor-isl} somewhat resembles the earlier proposed 
Ground State Computing
devices \cite{Lent,Bandy}. In both ideas the bistable
polarization of the basic cell as well as only the nearest neighbor
coupling are used. The main difference is the absence of the power
supply in Ground State Computing, so that the only driving force
is the fixed polarization of the cells at the ``edges'' of the device.
The small total energy gain (proportional to the number of ``edge'' 
cells) should be distributed evenly between all ``bulk'' cells to
ensure their deterministic sequential switching. Hence, an integrated
Ground State Computing device cannot operate in the mode of sequential
switching of cells. In order to reach the ground state, a significant 
part of the device should be involved in the macroscopic quantum 
process \cite{Aver-Odin,Aver-Naz} (``simultaneous'' switching of many
cells), and this transition would require practically infinite time
because of the exponential dependence on the number of cells.
	In contrast to the Ground State Computing, the principle of
operation of the Wireless Single Electron Logic allows the traditional
computing by the sequential switching of cells in the device of 
arbitrary large integration scale.

        There is no static power dissipation in Wireless Single Electron 
Logic. Typically the switching of a cell requires the energy of the 
order of only $e^2/C$ where $C$ is a typical capacitance. However, this
dissipation can be further reduced.
        In the recent suggestion called Single
Electron Parametron \cite{Param} the robust signal propagation 
along the shift register can cost even less, ultimately much less 
than $k_B T$ per switching of
a cell. The possibility of logic devices with the energy dissipation
below thermal limit was proven long ago \cite{belowKT,belowKT-2}. 
Single Electron Parametron seems to be the first realization of such 
a device based on the classical dynamics of 
the discrete internal degree of freedom. 

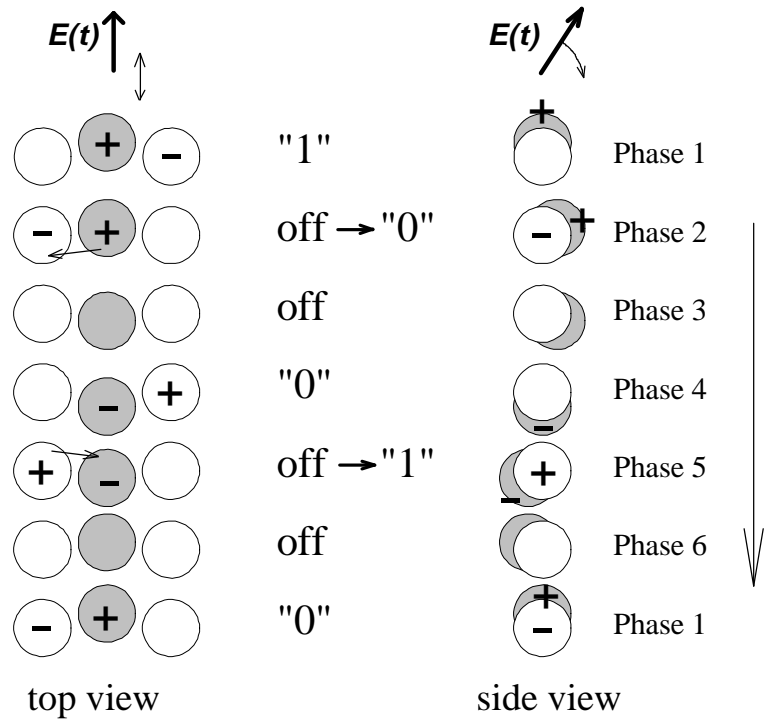
\begin{figure}
\caption{Shift register of the Single-Electron Parametron 
\protect\cite{Param}.
Information propagation is caused by the rotating electric field $E(t)$.}
\label{parametron}\end{figure}

        The idea is shown in Fig.\ \ref{parametron} (it represents the 
simplest, though not the best mode of operation). The basic cell is a chain
consisting of 3 islands. Rotating electric field changes the polarization
of the chain four times per period: islands are neutral (state ``off'') 
when the field is  perpendicular to the plane of the chain, and chain is 
necessarily polarized (state ``0'' or ``1'') when the field is in-plain. 
When the neutral state
becomes the polarized one, it can evolve into two different states:
the electron from the central island can jump on either of two outer
islands. The result is determined by the polarization of the neighboring
(previous) chain which has became polarized earlier because of the change
in the chain orientation along the propagation line. Notice that the 
next chain does not influence the decision because it is in a neutral
phase at this time. The resulting
polarization will in turn determine the polarization of the next chain
when it will enter the polarized phase.
        For the circuit shown in Fig.\ \ref{parametron} the signal 
propagation speed is  
6 steps per period of field rotation, and the transmission rate is 2 bits 
per period, so on average each bit requires 3 chains.

        The power dissipation less than $k_B T$ per switching of a chain
is achieved at low rotation frequency, $\omega << (k_B T)^2/e^3REd$,
where R is the tunnel resistance and $Ed$ is the voltage
between islands induced by the in-plane component of the field. In this case
the switching consists of the large number of electron jumps back and forth.
In the adiabatic limit the energy $k_B T \ln 2$ is first taken from the
thermostat (when neutral and polarized states have equal energies, the 
entropy is reduced by one bit) and then this energy is returned to the 
thermostat. In the first approximation the total power dissipation
per switching is proportional to the switching speed.

        In comparison with the Wireless logic of Ref.\ \cite{Kor-isl},
Single-Electron Parametron offers also larger parameter margins 
\cite{Param}.
       Numerical simulations for a particular ``layout'' show the 
margin about $20\%$ for the 
amplitude of the  rotating electric field.

        \subsection{ The problem of background charge}

        Fluctuating background charge is a very serious, possibly
the most serious problem of the integrated single-electronics.
Single-electron devices are so sensitive to the induced charge, that
a single charged impurity
in the close vicinity of a device can significantly influence its operation.
In the case of a single circuit, background charges can be adjusted 
individually with a help of additional gates. There is obviously no such      
possibility for integrated circuits. What could be a solution of
this problem?

        First, the problem may turn out to be not so serious after
all. There is some experimental evidence \cite{Lambe,Kuzmin-Likh} 
that even in rather dirty systems the
background charge tends to relax to zero. Theoretically this could be
understood, for example, as being due to the attraction of the
charged impurities to conducting surfaces by the image charge force.
In general, one can hope that the narrow statistical distribution of 
background charges might occur naturally in some materials.

        Second, it might be that the problem can be solved with the use of
extremely pure materials. For example, if we speak about molecular
electronic devices in which all circuit elements are reproducible on the
atomic level, we can imagine extremely low concentration of impurities.

        Third, instead of capacitively coupled single-electron devices
we can try to use resistively coupled circuits. For example, R-SET 
transistor is not influenced by background charges at all. However, 
there are problems along this way. The R-SET is obviously much more
difficult for fabrication than C-SET, and also
the R-SET as a voltage amplifier requires significantly 
lower temperatures \cite{R-SET} because of the Nyquist noise in
the coupling resistor. 

        Finally, one more possibility is to come up with some
capacitively coupled devices which would work in the environment
of fluctuating background charges. The particular example of
such a ``$Q_0$-independent device'' was suggested recently 
\cite{q0ind}.

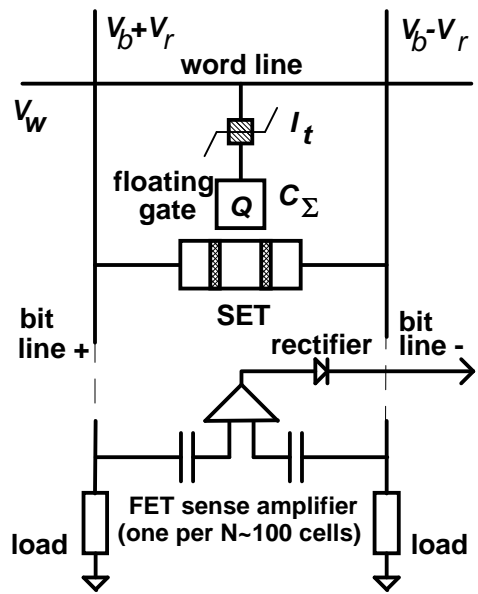
\begin{figure}
\caption{ Ultradense hybrid SET/FET memory operating in $Q_0$-independent
mode \protect\cite{q0ind}.}
\label{Q0-ind-fig}\end{figure}

        The idea is to use C-SET in a mode when the ramping input
signal drives the SET-transistor through several periods of its
control characteristic. In this case the output signal will
oscillate (Fig.\ \ref{SET-I-Q}), and for any initial $Q_0$ 
the amplitude of oscillation
is equal to the maximal swing of the control characteristic.
Such transistors can be used in a very-high density memory 
($10^{11}$ bits/cm$^2$ or even more) to read out the stored 
information \cite{q0ind} (Fig.\ \ref{Q0-ind-fig}). 
Suppose that similar to traditional
nonvolatile semiconductor memories \cite{flash-mem}, the digital bits 
are stored in a form of electric charge $Q$ on the 
floating gate located in the vicinity of SET-transistor.
In case of very small gate (of the order of 10 nm) this charge is
just a few (10--20) electrons. The charge can be changed, for example, 
by its injection/extraction through the dielectric layer via
Fowler-Nordheim tunneling \cite{flash-mem} (the graded barrier would
considerably improve the operation \cite{q0ind}). 
        The cell is selected by the simultaneous application of the 
voltages of different polarity to word and bit lines (small voltage
difference between two bit lines is used for the SET-transistor
biasing -- see Fig.\ \ref{Q0-ind-fig}). To read out
the stored information we try to write the logical unity in. If
unity has been already stored, the charge on the gate does not 
change, and the SET-transistor remains in the initial state. However,
if logical zero has been stored on the gate, then its charge will
gradually increase up to the level corresponding to logical unity.
During this increase the current through SET-transistor oscillates
that can be registered by a FET sense-amplifier (one FET may serve 
about 100 memory cells).
The previously stored information is destroyed during read-out, 
hence, it should be restored later.
    Notice that
voltage amplification by SET-transistor is not required in this mode
of operation, and this fact
significantly increases (by a factor $\sim$5) the maximal operation 
temperature.

        Estimates show \cite{q0ind} that the density of $10^{11}$
bits/cm$^2$ and the room temperature operation of such a memory
is feasible for $\sim$3 nm minimum feature size technology. Estimated
read/write time is about 3 ns and is limited both by the time of
the floating gate charging and by the intrinsic noise \cite{Kor-noise} 
of the SET-transistor.

        Single-electron devices operating in $Q_0$-independent
mode seem to be the most radical solution of the problem of the 
background charge fluctuations. However, despite this idea can
be used in the memory devices, it can be hardly 
applied to the logic circuits.

	\newpage
        \section[molecular]{Single-electronics in semiconductors,
        clusters of metal atoms, and molecular systems}

        Single-electron effects become stronger with the decrease
of the typical size. Besides that, they necessarily acquire new
features. Eventually the field of single-electronics transforms 
into the field of atomic physics and chemistry; however, the basic ideas 
of single-electron devices are applicable even at this level. They
can be used for the information processing in the hypothetical
molecular electronics devices.

        The ``orthodox'' theory works well for metallic systems
down to crudely 1-nm size scale. At this scale the level discreteness
in small metal particles (clusters of atoms) starts to play an 
important role. Besides that, increasing Coulomb
energy becomes comparable to the height of the tunnel barriers
leading to highly nonlinear $I-V$ curves.
        In semiconductors these effects are important at considerably
larger size scale and they are typical in experiments with quantum dots.
        Obviously, these effects should be also taken into account when the 
tunneling via single molecules is studied. That is why in this section
we consider together the features of single-electron circuits based on 
semiconductor quantum dots, clusters of metal atoms and single molecules.

        \subsection[quantization]{The level discreteness}

        The ``orthodox''  theory  of  single-electronics  assumes   the 
continuous
energy spectrum of all electrodes. It should be somewhat modified
\cite{Aver-Kor,Kor-dots,Beenakker} to take into account the level 
discreteness.
As an example consider the SET-transistor with discrete spectrum
of electrons in the central island. The complete description of 
the charge state 
now includes not only the total number of electrons on the island,
but also the occupation of individual levels. In one-electron 
approximation (neglecting the collective excitations \cite{Weinmann})
the electron addition energy depends on two integer parameters 
$k$ and $n$:
\begin{equation}
E_{k,n}=\varepsilon_k +e (ne+Q_0+e/2)/C_\Sigma,
\label{ekn}\end{equation}
        where $k$ is the level number ($\varepsilon_k$ is the energy
spectrum) and $n$ is the total number of excess electrons on the 
island.
	Notice that the contribution from the background charge can
be included into the definition of $\varepsilon_k$. The finite bias
voltage can be taken into account in the same way as for the usual
SET-transistor (Section IV).
 	The tunneling rates should be calculated for each level
individually. The rate of electron tunneling to/from the
empty/occupied $k$-th level via $j$-th junction is given by expression
\begin{equation}
\Gamma =\Gamma _j \, \frac{1}{1+\exp (-W/k_B T)} \, , \,\,\,
        W = \pm (-1)^j \varepsilon _k  \pm eV_j(n)-e^2/2C_\Sigma \, ,
\label{Gamma-discr}\end{equation}
        where $\pm$ stands for the direction of tunneling,
$V_j(n)$ is the voltage drop across the junction
given by Eq.\ (\ref{V_j}), and $\Gamma _j$ depends on the matrix
element of tunneling and electron density in the external electrode
($\Gamma _j$ can also depend on $n$ and $k$). For the calculation
of the average current and other characteristics Eq.\ (\ref{Gamma-discr})
should be supplemented by some model describing the energy relaxation
of the electrons on the island. Equations (\ref{rate}) and (\ref{W_j}) 
of the ``orthodox'' theory can be obtained by summing Eq.\ 
(\ref{Gamma-discr}) over all energy levels in the case of negligible
level spacing, Fermi-distribution of electrons on the island, and
constant $\Gamma_j$  (then $R_j=\delta /e^2 \Gamma_j$ where $\delta$
is the average level spacing).

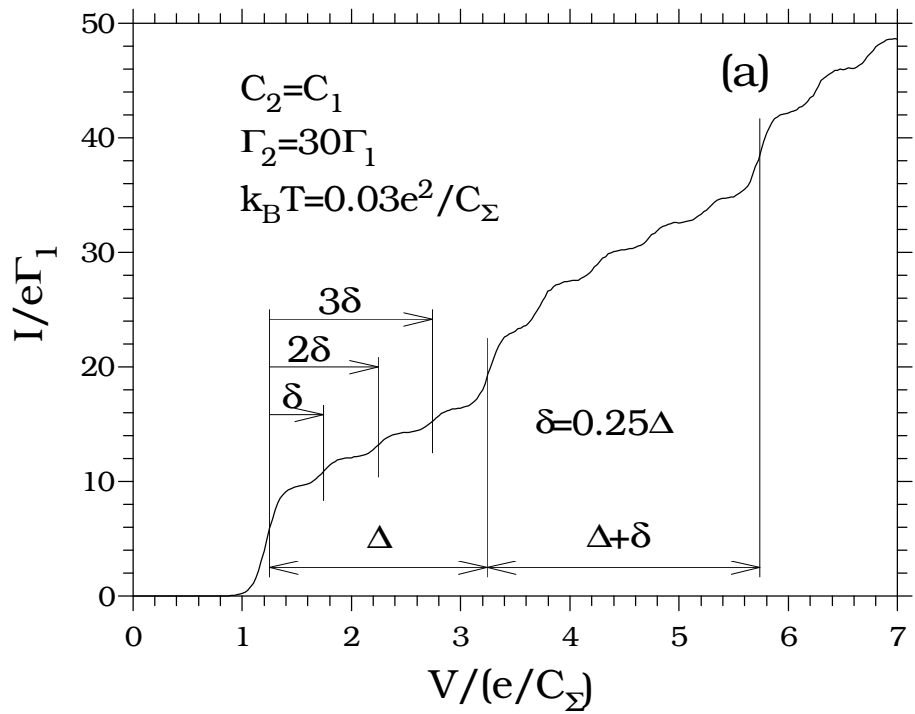
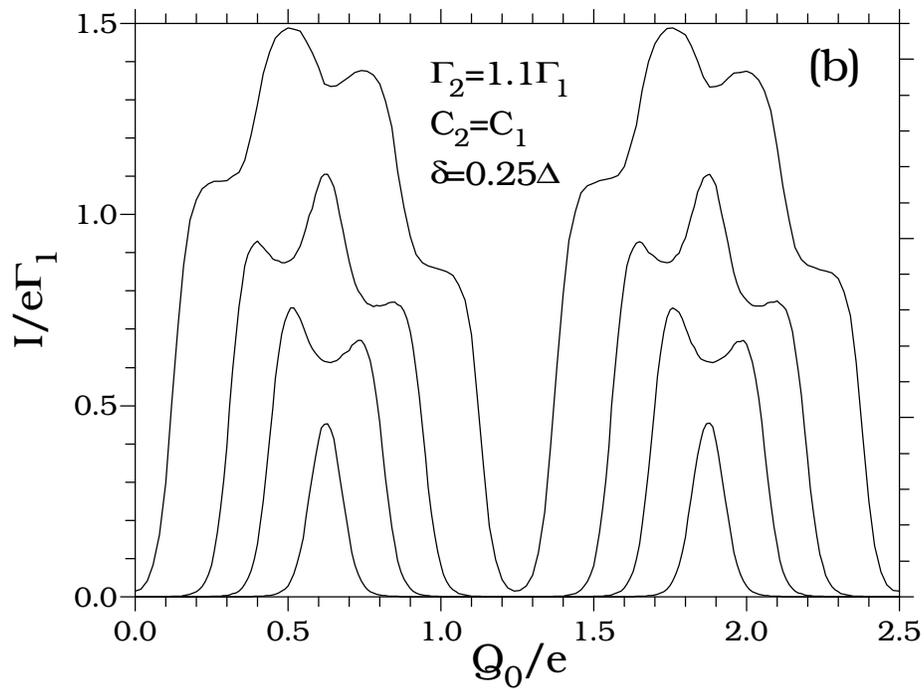
\begin{figure}
\caption{ (a) The typical $I-V$ curve and 
(b) the typical $I-Q_0$ dependence calculated for the 
SET-transistor with discrete spectrum of the central
island. Fine structure in (a) is due to the level
spacing $\delta$ while the Coulomb staircase is determined mainly
by the Coulomb energy $\Delta =e^2/C_\Sigma$. The $I-Q_0$ dependence
can have the multi-peak shape in contrast to usual SET (see Fig.\
\protect\ref{SET-I-Q}). }
\label{discret}\end{figure}

       The $I-V$ curve of a SET-transistor with level discreteness 
contains the step-like features (Fig.\ \ref{discret}a)  which appear
when the discrete level in the island crosses the Fermi level in
external electrode. The position of the step
along the voltage axis corresponds to $W=0$ in Eq.\ (\ref{Gamma-discr})
and depends on two integer parameters $k$ and $n$ (in contrast to only
one parameter $n$ in usual SET-transistor) as well as on the junction
number $j$. In the general case the arrangement of steps can be quite
complicated, however, in typical cases the simple classification is 
possible. For example, Fig.\ \ref{discret}a shows the $I-V$ curve in the
case when the level spacing  $\delta$ (equidistant two-fold degenerate 
spectrum is 
assumed) is considerably less than the Coulomb energy $e^2/C_\Sigma$,
and the barrier transparencies are significantly different.
        The level discreteness produces the fine 
structure superimposed on the Coulomb staircase. Notice that the 
level spacing contributes to the period of the
Coulomb staircase, hence, there is no pure periodicity in the case
of realistic nonequidistant spectrum $\varepsilon_k$.
        The slow energy relaxation of the electrons on the island
leads to some smoothing of the Coulomb staircase \cite{Aver-Kor}.
    
        The discrete levels also modify the dependence of the
current on the induced charge $Q_0$. In this case it can have the
multi-peak shape (compare Figs.\ \ref{discret}b and \ref{SET-I-Q}).
      The slight asymmetry of the peaks in Fig. \ref{discret}b is due 
to small difference between two tunnel barriers.
The perfect periodicity is absent if the spectrum $\varepsilon_k$
is not equidistant because it influences the position of the
peaks. The level spacing $\delta$ contributes also to the average
period making it larger than $e$.

        In the ``orthodox'' theory the total number of conducting 
electrons on the island is large, so that it is possible to extract
any number of them. This leads to some sort of electron-``hole''
symmetry. In the case of quantum dots or molecular-scale devices
it is possible to have just few conducting electrons on the island,
so that there is obviously no such a symmetry. Even the complete
asymmetry when initially there are no conducting electrons on the 
island and they appear only due to transport, is quite typical
(in semiconductors the same situation is also possible for holes).
In this case  the relative importance of the charge quantization and
energy discreteness depends not only on the ratio $\delta /(e^2/C_\Sigma )$
but also on the ratio $\Gamma_e/\Gamma_c$ of emitter and collector 
barrier transparencies  so that the actual parameter
is $\alpha=\delta /(e/C_\Sigma) \times \Gamma_e/(\Gamma_e+\Gamma_c)$
\cite{Kor-dots}. For example, even if $(e/C_\Sigma) > \delta$ but
the collector barrier is much lower so that $\alpha << 1$, than electrons do 
not accumulate on the central island, and steps on the $I-V$ curve
reflect only the spectrum $\varepsilon _k$. Coulomb staircase is
noticeable only when $\alpha \agt 1$.

        Equations (\ref{ekn}) and (\ref{Gamma-discr}) are based on the
classical expression $E_{int}=(me)^2/2C_\Sigma$ for the interaction 
energy of $m$ electrons on the island. In the few-electron case
(small $m$) the absence of the electrostatic self-interaction of an
individual electron makes this simple expression considerably 
inaccurate, and the better approximation is $E_{int}=m(m-1)e^2/
2C_\Sigma$ \cite{Aver-Kor}. The good accuracy of this approximation
is confirmed by exact calculation \cite{Belkhir} of the interaction
energy of few (up to 30) electrons on the sphere even in the
extremely ``quantum'' case when electrostatic energy is much smaller
than the Fermi energy.

        The separation of the electrostatic and one-electron energy
in Eq.\ (\ref{ekn}) is definitely only a simple approximation, and
in the exact theory the  many-body problem should be solved.
This problem is simplified in the case when only the low-temperature 
low-voltage conductance is studied, then the transport is determined
by the ground states of the configurations with $m$ and $m+1$ electrons
on the island. The finite-voltage case requires also the calculation of 
excitations. There is some progress in this direction (see, e.g.,
Refs.\ \cite{Leburton,Weinmann}). However, the exact calculation
is difficult not only because of the mathematical complexity of the 
problem, but also because the result is very sensitive to the geometry
of the island  which is usually not known accurately.
The most widely used approximation is still Eq.\ (\ref{ekn}), and it
surprisingly well explains the experimental data (in some experiments 
the slow variation of the capacitance should be also taken into account
-- see next subsection). 
        
        The theory of single-electron transport in systems with
discrete levels \cite{Aver-Kor,Kor-dots,Beenakker} was well confirmed
experimentally both in metal and semiconductor structures. Let us
discuss the difference of the typical parameters of these structures.
        First, let us estimate the energy level discreteness in a 
spherical cluster of aluminum atoms with
diameter $d=1$ nm (it would contain only about 30 atoms). In the free 
electron gas approximation the average
spacing $\delta$ (per spin) between levels is given by expression
\begin{equation}
\delta = \frac{1}{g(\varepsilon_F)v}= \frac{2 \hbar ^2 \pi ^2}
{ vm(3\pi^2\rho )^{1/3} } \, ,
\label{delta}\end{equation}
        where $v$ is the volume, $m$ is the effective electron mass, 
and $\rho$ is the electron concentration. For the table value
$\rho =1.8\times 10^{23}$ 
cm$^{-3}$ we obtain $\delta \approx$ 0.15 eV. Estimating  the
typical single-electron Coulomb energy, $\Delta =e^2/C_\Sigma$, let us take
$C_\Sigma = \beta 2\pi \varepsilon \varepsilon_0 d$ with $\varepsilon 
\approx 5$ and the geometrical factor $\beta \approx 3$; 
then $E_c \approx$ 0.2 eV. 
        We see that in metallic systems the level discreteness becomes
comparable to the Coulomb energy roughly at 1-nm size scale, and the 
influence of the energy quantization is negligible when the typical
size is larger than few nanometers. That is why the level discreteness
is so difficult to observe in metallic single-electron devices.

        The interplay between two effects in the metallic system 
was demonstrated
experimentally for the first time only recently \cite{Ralph} using 
the transport through a very small aluminum particle with volume about 
130 nm$^3$. The corresponding spacing was $\delta \approx 0.7$ meV while
the charging energy was $\Delta \approx 12$ meV (the geometry was close
to the plane capacitor that increased $C_\Sigma$ in comparison with the 
estimate above). Because of the relatively small energy scale, the level
discreteness showed up on the $I-V$ curve only at the temperatures below
2 K (most measurements were done at $T=0.3$ K), at larger temperatures
only Coulomb staircase was observed. Notice that the
step-like features for the aluminum electrodes in the normal state 
were transformed in this experiment into the peak-like features 
\cite{Ralph} for superconducting electrodes
because their shape directly corresponds to the density of states in 
electrodes.

        The step-like features due to the level discreteness superimposed
on the Coulomb staircase were also observed in the experiment \cite{Dubois}
with metal clusters Pt$_{309}$Phen$_{36}$O$_{30}$. The level 
spacing $\delta$ was up to  50 mV while the single-electron charging
energy was up to 500 mV, and the discreteness was clearly 
observed at 4.2 K (the measurements at higher temperature were not
reported in the paper). It is remarkable that the conducting particle 
used in this experiment can be described by the chemical formula
(hence, formally this is a single molecule), and the ``orthodox''
theory (modified for the account of discreteness) is still very well 
applicable to this system. 
	The experiments confirm 
that in metal systems the level 
discreteness is a small effect in comparison with single-electron 
charging effects when the size scale is larger than roughly 1 nm.

       In the semiconductor systems the level discreteness becomes
important at considerably larger size scale. This is caused by
typically much lower electron concentration 
and lower effective electron mass (see Eq.\ (\ref{delta})).
For example, in Si-based systems with doping level $\rho \sim
10^{21}$ cm$^{-3}$, $\delta$ would become comparable
to the Coulomb energy at $d \sim$ 5 nm. For much lower doping 
concentration the interplay between the level discreteness and
the Coulomb effects was reported \cite{Chou} at $d\sim$20 nm.
	The irregular position of the Coulomb oscillation peaks
in Fig.\ \ref{Chou-fig} \cite{Chou} can be ascribed to the irregular
energy difference between neighboring discrete levels. Fluctuations
of the peak height can be caused by the different tunneling matrix 
elements for different levels.

        A more dramatic increase of the level spacing occurs in 
semiconductor systems
with two-dimensional electron gas. In this case $\delta$ does not
depend on the electron concentration, $\delta=\pi \hbar ^2/2mS$,
where $S$ is the island area. Let us estimate electrical capacitance
of the conducting island of 2D gas as $C_\Sigma=\varepsilon 
\varepsilon_0 S/a$.
Here $a$ is the effective distance from a conducting electrode in the
plain capacitor geometry which is a good approximation when the
``vertical'' transport via quantum dot is studied. In the case
of ``lateral'' transport (when conducting electrodes are in the same
plain) this expression can be used with  $a\sim 0.2 d$ 
proportional to the diameter $d$ of the dot.
The ratio $\delta/(e^2/C_\Sigma)$ is equal to $\pi \hbar^2
\varepsilon \varepsilon_0/2ma=a_B/2a$ where $a_B=4\pi 
\varepsilon\varepsilon_0
\hbar ^2/me^2$ is the Bohr radius in the given material. That is a 
natural result since
by definition the Bohr radius corresponds to the length scale at which
Coulomb and quantum energies coincide. In GaAs the Bohr radius is
as large as 10 nm. This is why both the level discreteness
and the single-electron effects are important
 \cite{Groshev,Kor-dots,Beenakker} in experiments with electron 
transport through GaAs-based quantum dots 
\cite{Goldman,McEuen,Vaart,Klein,Gueret,Tewordt,Johnson,Foxman,Weis} 
when the size scale $a$ is comparable to 10 nm.
 Stressing the analogy with atomic
physics in which the Bohr radius determines the size of the electron 
orbit, semiconductor quantum dots are sometimes called ``artificial
atoms'' \cite{Kastner}.

        The relative importance of two effects is quite different
in experiments with the vertical and lateral transport via quantum
dots. In the vertical geometry $a$ is close to the barrier width
(in fact, the finite well width and the existence of two barriers
should be taken into account \cite{Kor-dots}). The typical barrier 
width is 3--10 nm. Hence, the ratio $\delta /(e/C_\Sigma )$ 
is typically on the order of unity, and even can be larger than unity.
That is why the level discreteness is always important in the vertical
transport via GaAs quantum dots and can be a major effect. In the first 
experimental study \cite{Reed} of such a transport there was no sign 
of the single-electron
charging effect, and the steps on the $I-V$ curve were determined purely
by the energy spectrum $\varepsilon _k$. That was because the collector
barrier was much more transparent than the emitter barrier leading 
to $\alpha << 1$ (see discussion above). Similar experiment 
\cite{Goldman} with the increased thickness of the collector barrier
showed the Coulomb staircase with a fine structure due to $\varepsilon _k$.
It was possible to change the major effect simply applying the different 
polarity of the voltage \cite{Goldman} because that interchanged the
emitter and collector. The interplay between two effects in the vertical
tunneling via quantum dot was also reported by several other groups
(see, e.g., Refs.\ \cite{Gueret,Tewordt}). 

        In experiments with the lateral transport via a quantum dot,
the single--electron charging energy is typically considerably larger
than the level spacing. The estimate above gives for GaAs dot $\delta /
(e^2/C_\Sigma) \sim 25 \mbox{nm} /d$, so that for the typical dot 
diameter $d \sim 0.5 \mu$m this ratio is about 0.05 (this ratio somewhat
increases if we take into account the capacitance increase due to 
the coupling to electrodes). Notice that the application of strong 
magnetic field changes the energy spectrum and can considerably
increase the level spacing $\delta$. 

        There were many experiments demonstrating the coexistence
of the energy and charge quantizations in the lateral transport via
a quantum dot. Let us mention several of them 
\cite{McEuen,Vaart,Klein,Johnson,Foxman,Weis} and the recent review
\cite{Kouwenhoven-rev} on this topic.  Figure \ref{discret-exp}a 
shows the experimental $I-V$ curve with the Coulomb staircase
and the fine structure due to the level discreteness \cite{Johnson}.
The inset shows the layout of the metal gates which form the quantum
dot in the two-dimensional electron gas beneath them. The dependence
of the current on the voltage of the central gate C is shown in
Fig.\ \ref{discret-exp}b. This gate does not affect much the tunnel
barriers but changes the induced charge in the dot. The multi-peak
shape of the dependence is the consequence of the level discreteness
(compare with Fig.\ \ref{discret}b).

        The theory described above can be also applied to the
tunneling through single molecules. Experimental $I-V$ curves
in such systems \cite{Nejoh1,Nejoh2,Fischer,Zubilov} typically have 
a region of Coulomb blockade
and cusps or steps resembling Coulomb staircase, and these 
features are usually discussed in terms of single-electron
transport.
        If the molecule contains the relatively large cluster
of metal atoms (see, e.g., Ref.\ \cite{Dubois}), the good agreement
even with the simple ``orthodox'' theory can be expected. However,
if the cluster consists of just few atoms or there is no metal cluster
at all, the theory of single-electronics should be used with some 
caution.
         First   of   all,   the   level discreteness
is not a small correction in this case but a major factor.
Typically the separation of the Coulomb energy and one-electron
spectrum assumed in Eq.\ (\ref{ekn}) should fail, and the excitation
spectrum should considerably depend on the charge number $n$. 
The calculation of capacitances could be used only for crude estimates
and typically there should be no symmetry between addition and removal
of electrons (electron affinity and ionization energy can be quite 
different) leading to highly asymmetric $I-V$ curves.
In contrast to ``orthodox'' theory, it can be 
impossible to add or remove more than 2--3 electrons to/from
a molecule without its mechanical breakdown or chemical transformation.
        Thus, the experimental results in the single-molecule
systems can considerably differ from the predictions of the 
standard theory of single-electronics.

        On the other hand, it is surprising that such macroscopic
quantity as the capacitance can be sometimes used even at the 
microscopic size scale. In the model considered in Ref.\ 
\cite{Belkhir} only a few conducting electrons are sufficient
to establish a well-defined capacitance (the formal definition
fluctuate only slightly with the number of electrons). As a curious
example let us mention that the first three ionization energies of the
single aluminum atom (5.97 eV, 18.8 eV, and 28.5 eV \cite{LB}) correspond 
to the dimensionless sequence 1 : 3.1 : 4.8 which is very close to the 
``orthodox'' sequence 1 : 3 : 5.

        \subsection{Barrier dependence on the voltage}

        In the previous subsection the single-electron charging energy
$e^2/C$
of the 1-nm aluminum grain was estimated as 0.2 eV. This number can be
comparable to the energy height $H$ of the tunnel barrier which depends 
on the material and is typically between 0.3 eV (thermally grown
aluminum oxide) and 3 eV (vacuum barrier). This would lead to highly
nonlinear $I-V$ curves of single-electron devices.

        If the barrier has low transparency then the typical
voltage of the $I-V$ curve nonlinearity is even much less than $H/e$
and is comparable to $\hbar /e\tau$ where $\tau$
is the traversal time of tunneling (in case of the rectangular barrier
$\tau =l/(2H/m)^{1/2}$ where $l$ is the barrier width). For example,
if $H=1$ eV and $l=2$ nm, then $\hbar/e\tau$=0.2 eV.

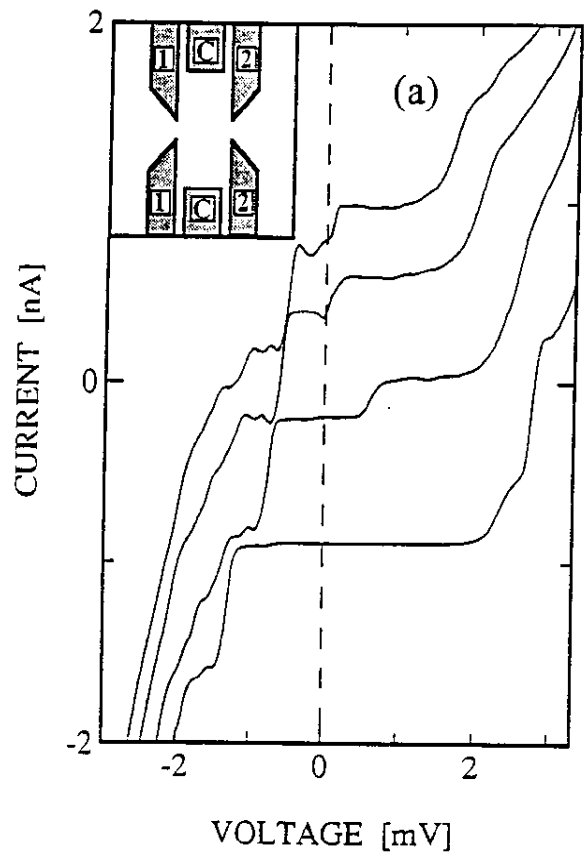
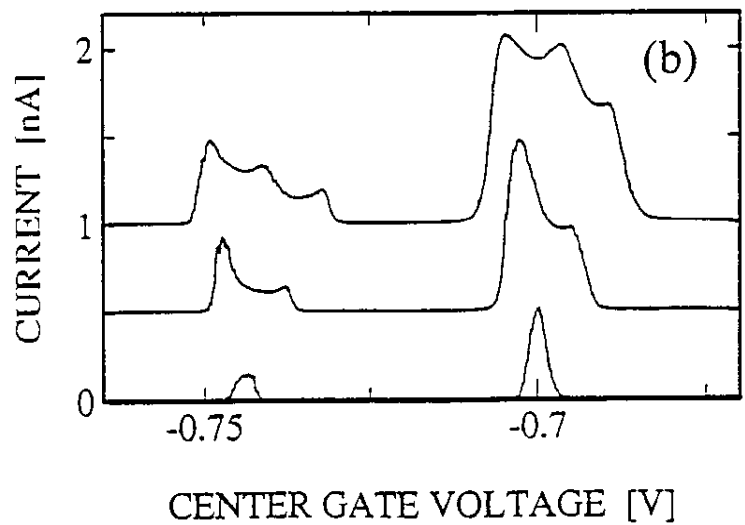
\begin{figure}
\caption{ (a) The experimental $I-V$ curve and (b) the dependence
of the current on the gate
voltage \protect\cite{Johnson} for the C-SET based on the GaAs quantum
dot (courtesy of A.\ T.\ Johnson). Notice the fine structure on the $I-V$ 
curve and the multi-peak shape of curves
in (b) due to the level discreteness. Also notice the nonlinearity of
the $I-V$ curve because of the barrier suppression.}
\label{discret-exp} \end{figure}

        Hence, the finite height of the barrier becomes an
important factor \cite{Av-Likh-mes,Kor-Naz,Wada} in metallic
single-electron devices typically at the size scale of 1 nm
(in some materials it appears considerably earlier \cite{Pashkin-Cr}).
        The suppression of the tunnel barrier by the applied voltage 
is always the strong effect in experiments with lateral transport 
via semiconductor quantum dots because of typically low barrier
height. For example, Fig.\ \ref{discret-exp}a shows the experimental 
$I-V$ curves \cite{Johnson} in which the voltage scale of the exponential 
nonlinearity of the $I-V$ curve is comparable to the period of the
Coulomb staircase. Notice that in semiconductor devices the barrier 
suppression is typically important even when the relatively large size
scale does not allow to resolve individual levels.
         Finite barrier height is obviously also important
in the single-molecule systems because of large typical voltages.
 
        The effect can be taken into account within ``orthodox''
theory (neglecting for simplicity the level discreteness) by introduction
of the nonlinear ``seed'' $I-V$ curve $I_0(V)$ of the tunnel junction
\cite{Av-Likh-mes}.
The tunneling rates in this case are given by the general expression
\begin{equation}
\Gamma = \frac{I_0(W/e)}{e(1-\exp (-W/k_B T))}
\label{Gamma-nl}\end{equation}
instead of Eq.\ (\ref{rate}) ($W$ is the energy gain due to tunneling).
A more accurate approximation \cite{Kor-image,Av-image} takes into 
account the change of the image charge potential due to Coulomb
blockade and gives  the additional factor $\exp (e^2\tau / 12C\hbar)$.

        When the nonlinearity of $I_0(V)$ is relatively small at the 
single-electron voltage scale $V\sim e/C$ (it implies $\hbar /\tau 
<<e^2/C$), the $I-V$ curve of the SET-transistor preserves usual
Coulomb features. However, the current grows exponentially with voltage,
so that it becomes impossible to measure experimentally the offset voltage,
and the Coulomb staircase becomes smoother \cite{Kor-Naz}
(see Fig.\ \ref{discret-exp}a \cite{Johnson}). In case
of strong nonlinearity (which is not yet achieved experimentally)
Coulomb staircase should completely disappear and give place to the
new periodic features with different period \cite{Kor-Naz}.

        Let us mention one more effect which is important
in semiconductor single-electron devices. In contrast to metallic 
systems, the geometrical size of a  semiconductor conducting island 
can depend on
the number of electrons on the island and on the gate voltage. Hence,
the capacitance is not constant, leading to nonperiodicity of
the Coulomb staircase and the nonperiodic dependence on the gate voltage 
in SET-transistor. The change of the geometric size also leads to 
the change of the width of a tunnel barrier while the barrier height 
can be directly affected by the gate voltage. Sufficiently large
gate voltage can either completely deplete the conducting island or
remove  the  tunnel  barrier  depending  on  the  polarity.  As   a 
consequence, on the large scale of the gate voltage
semiconductor SET-transistors usually behave like FET transistors
(see Figs.\ \ref{Chou-fig} and \ref{discret-exp}b):
starting from the state with
negligible current, one can finish with the perfectly open transport
channel. Single-electron quasiperiodic dependence on the gate voltage
(Coulomb oscillations)
is observed in relatively narrow range of the gate voltage when the 
conducting island has already appeared but the tunnel resistance of 
the barrier is still larger than the quantum unit $R_Q$.

	\newpage
        \section[conclusion]{Conclusion}

        We have considered only a part of issues related to the field
of single-electronics. For example, we did not mention single-electron
effects in superconducting systems \cite{Av-Likh-mes},
including the possibility 
to measure experimentally the parity of the total number of electrons
in superconducting islands \cite{Tuominen,parity}. Another interesting 
subject is the single-electron oscillations with the frequency determined
by the dc current, $f=I/e$ 
\cite{Av-Likh-mes,Bakhvalov,Kuzmin-Bloch,Pashkin,Kor-Bl,Kor-oscil}.
This relation can be inverted: the magnitude of the dc current can be
accurately controlled by the frequency of applied ac bias 
\cite{Delsing,Geerligs,Pothier-pump} that is used in single-electron
turnstile \cite{Geerligs} and pump \cite{Pothier-pump}. We have not discussed
also the problem of cotunneling \cite{Aver-Odin,Aver-Naz},
the effect of the electromagnetic environment \cite{Nazarov,Nazarov-Ingold},
photon-assisted tunneling \cite{Likh-phot,Bruder,Leo-phot},
coherent effects \cite{Yacoby,Buttiker},  and many other issues.

        Single-electronics was a rapidly growing field during the last
ten years, and this growth still continues. It is already clear that
single-electronics is interesting not only from the scientific point 
of view, but it can be really used in applications. The simplest
application is the use of the SET-transistor for various purposes 
as a very sensitive electrometer capable to measure sub-electron charges.
Another clear application is the standard of dc current 
\cite{Martinis-pump} based on the single-electron pump. 
It is quite possible that arrays of small tunnel junctions will
be used as low-temperature thermometers \cite{Pekola-PRL}. Other
applications are definitely coming.

        The most important potential application is the
ultradense (up to 10$^{12}$ cells per cm$^2$) integrated digital 
electronics which was the main topic of the present review. 
The question if such a prospect is real, is still quite uncertain
because of very difficult problems on this way. The main problem
is the necessity of the new technology capable to deal with objects
on the order of 1 nm or even less. This length scale is imposed
by the requirement of the room-temperature operation. It is likely
that such a technology should use conducting clusters of atoms
embedded in the molecular matrix, hence, we speak about the 
molecular electronics devices. Another major obstacle on the way
to integrated single-electronics is the random distribution of the 
background charge. If the technology will not offer the solution of 
this problem, only circuits operating in $Q_0$-independent mode
\cite{q0ind}
will be practical, and this will considerably limit the variety
of possible devices.

        Despite the problems, the ultradense integrated
single-electronic circuits will hopefully be eventually realized and will
be able to substitute the CMOS technology to continue the exponential
growth of the computer performance. The rapid progress in experimental
single-electronics, in particular, the recent demonstration of the
devices operating at the temperature of liquid nitrogen and even at
room temperature, strongly supports this hope.

\vspace{1cm}

        The author thanks D.\ V.\ Averin and K.\ K.\ Likharev for the
numerous discussions and the critical reading of the manuscript.
The author is also grateful to Y.\ Nakamura, E.\ Leobandung,
P.\ D.\ Dresselhaus, and A.\ T.\ Johnson for providing figures 
with experimental results.
The work was supported in part by ONR grant No. N00014-93-1-0880 
and AFOSR grant No. 91-0445.

\end{document}